\documentclass[apj]{emulateapj}
\usepackage{url,amsfonts,epsfig}
\usepackage{graphicx}
\usepackage{amsmath}
\usepackage{amssymb}
\usepackage[section]{placeins}
\usepackage{chngpage}
\usepackage{calc}
\usepackage{subfigure}
\usepackage{mathtools}

\usepackage{natbib}
\makeatletter
\renewcommand\@biblabel[1]{}

\shorttitle{Formation of the Milky Way Nuclear Star Cluster}
\shortauthors{Antonini et al.}
\begin{document}
\def\gap{\;\rlap{\lower 2.5pt
\hbox{$\sim$}}\raise 1.5pt\hbox{$>$}\;}
\def\lap{\;\rlap{\lower 2.5pt
\hbox{$\sim$}}\raise 1.5pt\hbox{$<$}\;}
\newcommand\sbh{SMBH}
\title{Dissipationless Formation and Evolution of the Milky Way Nuclear Star Cluster}

\author{Fabio Antonini}
\email{antonini@cita.utoronto.ca}
\affil{Canadian Institute for Theoretical Astrophysics, University of Toronto,
60 St. George Street, Toronto, Ontario M5S 3H8, Canada}
\author{Roberto Capuzzo-Dolcetta}
\affil{Department of Physics, Sapienza-Universit\`{a} di Roma, P.le A. Moro 5,
I-00185, Rome, Italy}
\author{Alessandra Mastrobuono-Battisti}
\affil{Department of Physics, Sapienza-Universit\`{a} di Roma, P.le A. Moro 5,
I-00185, Rome, Italy}
\author{David Merritt}
\affil{Department of Physics and Center for Computational Relativity and Gravitation, Rochester Institute of Technology, 85 Lomb Memorial
Drive, Rochester, NY 14623, USA}

\begin{abstract}
In one widely discussed model for the formation of nuclear star clusters (NSCs),
massive globular clusters spiral into the center of a galaxy and merge to form
the nucleus.
It is now known that at least some NSCs coexist with supermassive black holes
(SBHs);
this is the case, for instance, in the Milky Way.
In this paper, we investigate how the presence of a \sbh\ at the center of the
Milky Way
impacts the merger hypothesis for the formation of its NSC.
Starting from a model consisting of  a low-density nuclear stellar disk and the \sbh,
we use direct $N$-body simulations to follow the successive inspiral and
merger
of globular clusters.
The clusters are started on circular orbits of radius 20 pc, and their initial
masses and radii are set up in such a way as to be consistent with the galactic tidal field
at that radius.
These clusters, decayed orbitally in the central region due to their large mass, were followed in their inspiral events; 
as a result, the total accumulated mass by $\approx 10$ clusters is about
$1.5\times 10^7M_\odot$.
Each cluster is disrupted by the \sbh\ at a distance of roughly one parsec.
The density profile that results after the final inspiral event is characterized
by a core
of roughly this radius, and an envelope with density that falls off $\rho\sim
r^{-2}$.
These properties are similar to those of the Milky Way NSC, with the exception
of the core size, which in the Milky Way is somewhat smaller.
But by continuing the evolution of the model after the final inspiral event, we
find
that the core shrinks substantially via gravitational encounters in a time
(when scaled to the Milky Way) of 10 Gyr as the stellar distribution evolves
toward
a Bahcall-Wolf cusp.
We also show that the luminosity function of the Milky Way NSC is consistent
with the
hypothesis that 1/2 of the mass comes from old ($\sim 10$ Gyr) stars,
brought in by globular clusters,
with the other half due to continuous star formation.
We conclude that a model in which a large fraction of the mass of the Milky Way
 is due to infalling globular clusters is consistent with existing observational constraints.
\end{abstract}

\section{Introduction}

The centers of low-luminosity spheroids, $M_B\gap -18$, are often
marked by the presence of compact stellar nuclei with half-light
radii of a few parsecs and luminosities that are $\sim 20$ times that of a
typical
globular cluster \citep{carollo,BSM:02,Cote}.
The nearest such system is at the center of our galaxy \citep{S011}.
The Milky Way NSC is close enough that its radial and kinematical structure
can be resolved \citep{ShEc,S09}.
Its total mass is estimated at $\sim 10^{7}M_{\odot}$ \citep{LZM,S08} and its half-light radius is roughly 3-5
pc \citep{GS:09,S09,S011}.
There is a central core of radius $\sim 0.5$ pc \citep{Buch},
beyond which the density falls off roughly as $r^{-1.8}$ \citep{Genzel03,ShEc,Oh}.

NSCs with properties similar to those of the Milky Way have now been
detected in galaxies of all Hubble types; a review of their properties
is given by \citet{BKR10}.
NSCs exhibit complex star formation histories; while the bulk of the
stars appear to always be old, the fraction of young stars increases
toward late-type galaxies.
In galaxies beyond the Local Group, NSCs are typically unresolved,
and the only structural properties that can be derived are half-light radii
and total luminosities.
The study of NSCs has  raised considerable interest because of the
fairly strong correlations between their masses and the properties
(mass, velocity dispersion) of their
host galaxies \citep{F:06,WH:06}.
These correlations suggest that the formation of NSCs and their host
galaxies are linked in important ways.

Two models for the formation of NSCs have been widely discussed.
In the in-situ formation model,  buildup of molecular gas near the center
of a  galaxy leads to episodic star formation events~\citep[e.g.,][]{LT82}.
In this model, a NSC consists mostly of stars that formed locally \citep{S06,Sch08}.
A number of mechanisms have been discussed for bringing the gas to the center,
including the magneto-rotational instability in a differentially rotating gas
disk \citep{Milos04},   tidal compression  
in shallow density profiles \citep{EV:08} or dynamical
instabilities  \citep{SB:89,B07}.

Alternatively, in the merger model,  globular clusters sink to the center of a
galaxy
via dynamical friction and merge to form a compact stellar system
\citep{TOS,CD93,AM:11}.  Observations of NSCs in dwarf elliptical
galaxies
suggest that the majority of such nuclei might have formed in this way \citep{lotz}.  
Numerical simulations have also shown that the basic properties of  NSCs are
consistent with a merger origin \citep{B04,CDM08,Hartmann}.

In addition to a NSC, the Milky Way also contains a massive black hole (\sbh)
whose mass, $M_\bullet\approx 4\times 10^6 M_\odot$  
\citep{Ghez08,Gil},
is comparable with that of the NSC.
A handful of other galaxies are also known to contain both a NSC and a \sbh\
\citep{seth,GS:09}, and the
ratio of \sbh\ to NSC mass in these galaxies is of order unity.
The possibility of a direct link  between the population of intermediate mass black holes that might form 
 in orbitally decaying star clusters and the growth of supermassive black holes at the center of galaxies has 
 also been suggested in previous papers~\citep[e.g.,][]{EBIS,PZ06}.

A simple argument leads to a $1$~pc scale as the relevant one for the  merger model for the formation of NSCs.
The beginning of the disruption process of a globular cluster due to tidal stresses from a \sbh\ is expected
when it passes within a certain distance of the galaxy center,
limiting the density within that radius.
Disruption occurs at a distance $r=r_\mathrm{disr}$ from the \sbh, where
\begin{equation}\label{eq:rtidal}
\frac{M_\bullet}{\frac43\pi r_\mathrm{disr}^3} \approx \rho(0)
\approx \frac{9}{4\pi G}\frac{\sigma_K^2}{r_K^2} .
\end{equation}
Here $\rho(0)$ is the central (core) density of the globular cluster,
$\sigma_K$ its
central, one-dimensional velocity dispersion, and $r_K$ its core radius;
the second relation is the ``core-fitting formula'' \citep{K66}.
Writing
\begin{equation}\label{rinfl}
r_\mathrm{infl} \equiv \frac{GM_\bullet}{\sigma^2_\mathrm{NSC}}
\approx 1.3 \mathrm{pc} \left(\frac{M_\bullet}{4\times 10^6M_\odot}\right)
\left(\frac{\sigma_\mathrm{NSC}}{100 \mathrm{km\ s}^{-1}}\right)^{-2}
\end{equation}
for the gravitational influence radius of the \sbh, where $\sigma_\mathrm{NSC}$
is the
stellar velocity dispersion in the NSC, equation~(\ref{eq:rtidal}) becomes
\begin{equation}
r_\mathrm{disr} \approx 2
\left(\frac{\sigma_\mathrm{NSC}}{5\sigma_K}\right)^{2/3}
\left(\frac{r_\mathrm{infl}}{r_K}\right)^{1/3} r_K.
\end{equation}
Setting $r_K=0.5$~pc and $\sigma_K=20$~km~s$^{-1}$, values characteristic
of the most massive globular clusters, we find $r_\mathrm{disr}\approx 1$~pc for
the Milky Way.
This is roughly equal to the radius of the core ($\sim 0.5$ pc) that is observed
in the distribution of late-type stars \citep{Buch}.

In this paper, we use direct  $N$-body simulations to test the merger
model for the formation of the Milky Way NSC.
Our initial conditions consist of a \sbh\ and a diffuse stellar component that
models
the inner parts of the nuclear stellar disk \citep{LZM}.
The NSC is  built up by the successive inspiral of globular clusters, which
we inject into the system at a radius of 20 pc.
The clusters are assigned masses and radii consistent with those of globular
clusters that were initially very massive ($\sim 4\times 10^6M_\odot$) but
which were tidally limited by the Galaxy's tidal field.
As the clusters spiral in due to dynamical friction against the stars in the
disk component, they eventually come within the radius of tidal disruption of
the \sbh.
We follow 12 such inspirals, resulting in the accumulation of $\sim
10^7M_\odot$.
The NSC that results has properties that are consistent with the observed
properties of the Milky Way NSC, including a $\rho\sim r^{-2}$ density profile
and a parsec-scale core.

At the Galactic center, the  relaxation time at SgrA*'s influence radius is
roughly
$20-30$ Gyr, assuming Solar-mass stars \citep{Mer10}.
This is too long for a Bahcall-Wolf~(1976) cusp to have formed over the Galaxy's
lifetime,
consistent with the observed lack of a cusp  \citep{Buch,Do,Bartko}.
But a pre-existing core with radius smaller than the \sbh\ influence radius would
have shrunk appreciably over a time of 10 Gyr due to gravitational encounters \citep{Mer10}.
We investigate the effect of such evolution on our NSC model by continuing
the $N$-body integrations after the final infall event, for a time that
corresponds
to roughly $10$~Gyr after scaling to the Milky Way.
The core radius decreases by roughly a factor of two in this time, bringing it
to a size that is more consistent with the observed core size.
The density profile beyond the core remains nearly unchanged.

Since Galactic globular clusters are almost exclusively ancient objects with typical ages$~10-13~$Gyr~\citep[e.g.,][]{Ros},
 the merger model  predicts that the bulk of the nuclear population is in old stars.
Accordingly, using   {\it Hubble Space Telescope}  Near-Infrared Camera and
Multiobject spectrometer~(NICMOS)  imaging of the inner $30~$pc of the Galaxy, we show that
the luminosity function of the Milky Way NSC is consistent
with the hypothesis that a large fraction of its mass is in ancient stars.

The paper is organized as follows.
The details of our initial models are given in Section~2.
Section~3 describes our simulations and results.
Section~4 is devoted to the study of the collisional evolution of the NSC following
its formation. The
implications of our results in the contest of  NSCs and Galactic center dynamics
are discussed in  
Section~5. Section~6 sums up.

\section{Initial conditions }
\begin{figure*}
\centering
{\includegraphics[width=0.8\textwidth ]{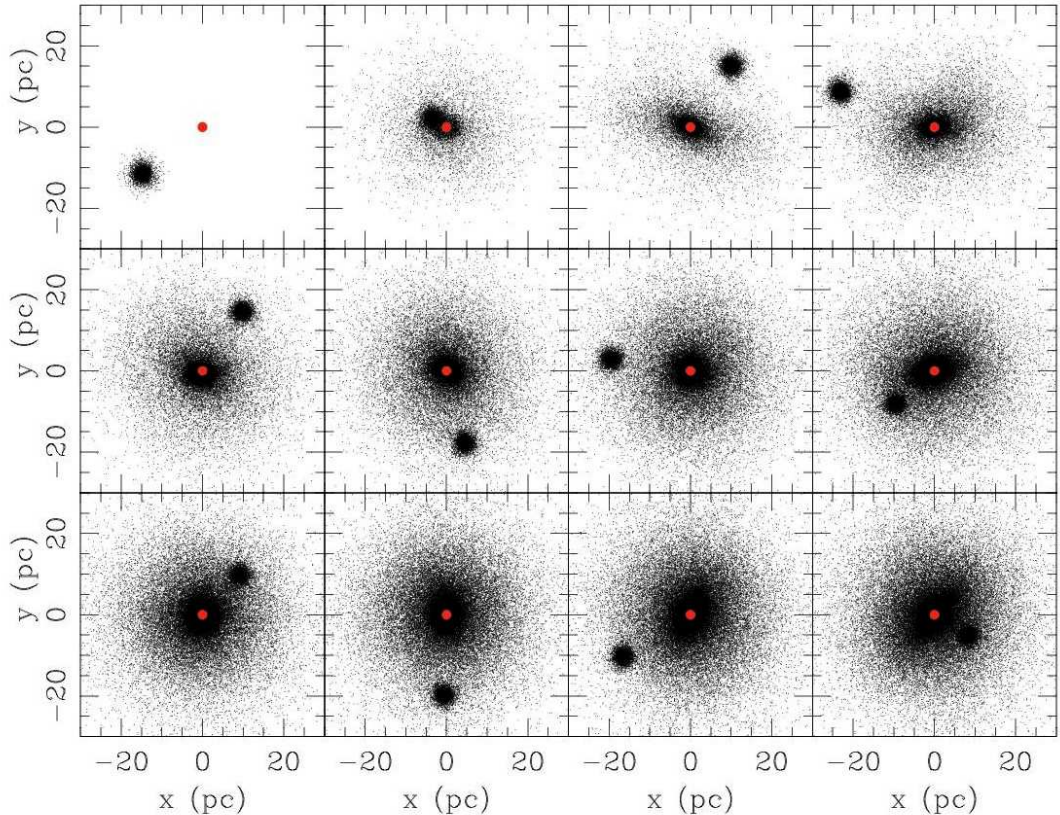}  }
\caption{Snapshots from the $N$-body integrations, projected onto a fixed
($x-y$) plane at the start of each infall event.
Only particles coming originally from the infalling clusters are displayed.
The \sbh\ is shown as the red circle.
}\label{snap}
\end{figure*}

In the following, we describe direct N-body simulations that were used to study  the  consecutive infall and merging
of a set of 12 globular clusters each starting from a galactocentric distance of $20$~pc.
The total mass of the 12, tidally-truncated clusters sums to $\sim 1.5\times 10^7M_{\odot}$, roughly the observed mass of the Milky Way nuclear star cluster~\citep{S09}.  We chose 12 because this number  permits a particularly elegant algorithm for 
generating the orbital initial conditions that does not favor any particular direction, as described in Section 2.2. 
After the first cluster had spiraled in to the center and reached a nearly steady state (as evaluated via Lagrange radii), 
we added the second cluster. 
This procedure was iterated until 12 clusters  accumulated and merged in the inner region of the galaxy
where we initially placed a central \sbh. Snapshots from the simulations are given in Figure~\ref{snap}.
This scheme  has been adopted to represent a realistic frame where the time interval between two consecutive globular cluster
 infalls and mergers to the center is longer than the time required for a single globular cluster 
 to reach a quasi-steady state following its merger into the growing NSC.
 In any case, the results of the following simulations are robust and not depend (as far as the structure of the emerging NSC is concerned) much on specific initial conditions, as we discuss at the end of Subsection~\ref{densityp}.

We now outline   the details of  the initial conditions adopted in the simulations.

\subsection{The Galactic Model}
The nuclear bulge  is distinguished from the larger Galactic bulge (effective radius $\sim 1~{\rm kpc}$)
by its flat disk-like morphology,  high stellar densities, and a history of  continuous   star formation.
The nuclear bulge dominates the inner $300~{\rm pc}$ of the Milky Way and it appears as  a, distinct, massive
disk-like complex of stars and molecular clouds which is, on a large scale,
symmetric with respect to the Galactic center.
It consists of an $r^{-2}$ nuclear stellar cluster within the inner $\sim 30~$pc, a larger nuclear
stellar disk  and a  nuclear molecular disk of same size (radius $\sim 200~{\rm pc}$ and scale height $\sim 45~{\rm pc}$).
The total stellar mass and luminosity of the nuclear bulge are $1.5\times 10^9~\rm M_\odot$ and $\sim2.5\times10^9~{\rm L_{\odot}}$,
respectively \citep{LZM}.
The   $r^{-2}$ density distribution
holds only  within the NSC in the central $\sim 30~{\rm pc}$, while, at larger radii,
the mass distribution is dominated by the nuclear stellar
disk which has essentially a flat density profile \citep{S011}.
The initial conditions for the galaxy in our simulations  model the nuclear stellar disk
and they omit the  central NSC. 
Accordingly, they correspond to a shallow density cusp around a
\sbh, which is included as a massive particle,
$M_{\bullet}=4\times10^{6}$M$_{\odot}$,
located at the origin.

We  adopted a truncated power-law model for this component:
\begin{equation}\label{dm}
\rho_{gx}(r)=\tilde{\rho}\left(\frac{r}{\tilde{r}}\right)^{-\gamma}\text{sech}\left(\frac{r}{r_{\rm cut}}\right).
\end{equation}
where $\tilde{\rho}=400$M$_{\odot}/$pc$^{3}$  is the density at  $r=\tilde{r}=10$pc,
and the truncation function is the same used by \citet{MD05}.
Since $\text{sech}(x) \approx 1-\frac{x^{2}}{2} $ for $x\ll 1$, the model is essentially a
power law at $r\ll r_\mathrm{cut}$, but it tends exponentially to zero for $r\gg r_{\rm cut}$.
We chose $\gamma =0.5$, corresponding to the  shallowest power law  consistent with an   isotropic velocity distribution  in  a point-mass potential.
The resulting model implies a mass density at $10~{\rm pc}$ similar to
what is found in the Galaxy outside the NSC ($\sim 400 ~M_{\odot}/{\rm {pc^{3}}}$).
We chose $r_{\rm cut}=22~$pc which gives a total mass of the (truncated) galactic model equal to $9.1\times10^7\text{M}_\odot$.

In order to generate a Monte-Carlo realization of the distribution function corresponding to the truncated density profile of equation (\ref{dm})
we followed the method described in \citet{SM05}.
Using Eddington's formula, it can be shown that the cumulative 
fraction of stars at radius $r$ with
velocities less than  $v$ is:
\begin{eqnarray}\label{fvr}
\centering
F(<v,r) &=& 1 - {1\over \rho}\int_{0}^E d\phi' {d\rho\over d\phi'}  \times  \\
&&\left\{ 1 + {2\over\pi} \left[{v/\sqrt{2}\over\sqrt{\phi'-E}} - \tan^{-1}\left({v/\sqrt{2}\over\sqrt{\phi'-E}}\right)\right]\right\},~~~\nonumber
\end{eqnarray}
where $E=\frac{1}{2}v^2+\phi(r)$ and  $\phi(r)$  is the total gravitational potential produced by the
stars and the \sbh. Once the  positions are assigned, 
equation (\ref{fvr}) can be numerically solved to  distribute
the particles in  velocity space.

The number of particles used to represent the galaxy was $N=240,000$, which implies a
 mass of $\sim 380\text{M}_\odot$  for each particle in the system.

\begin{figure}
\centering
{\includegraphics[width=0.36\textwidth ]{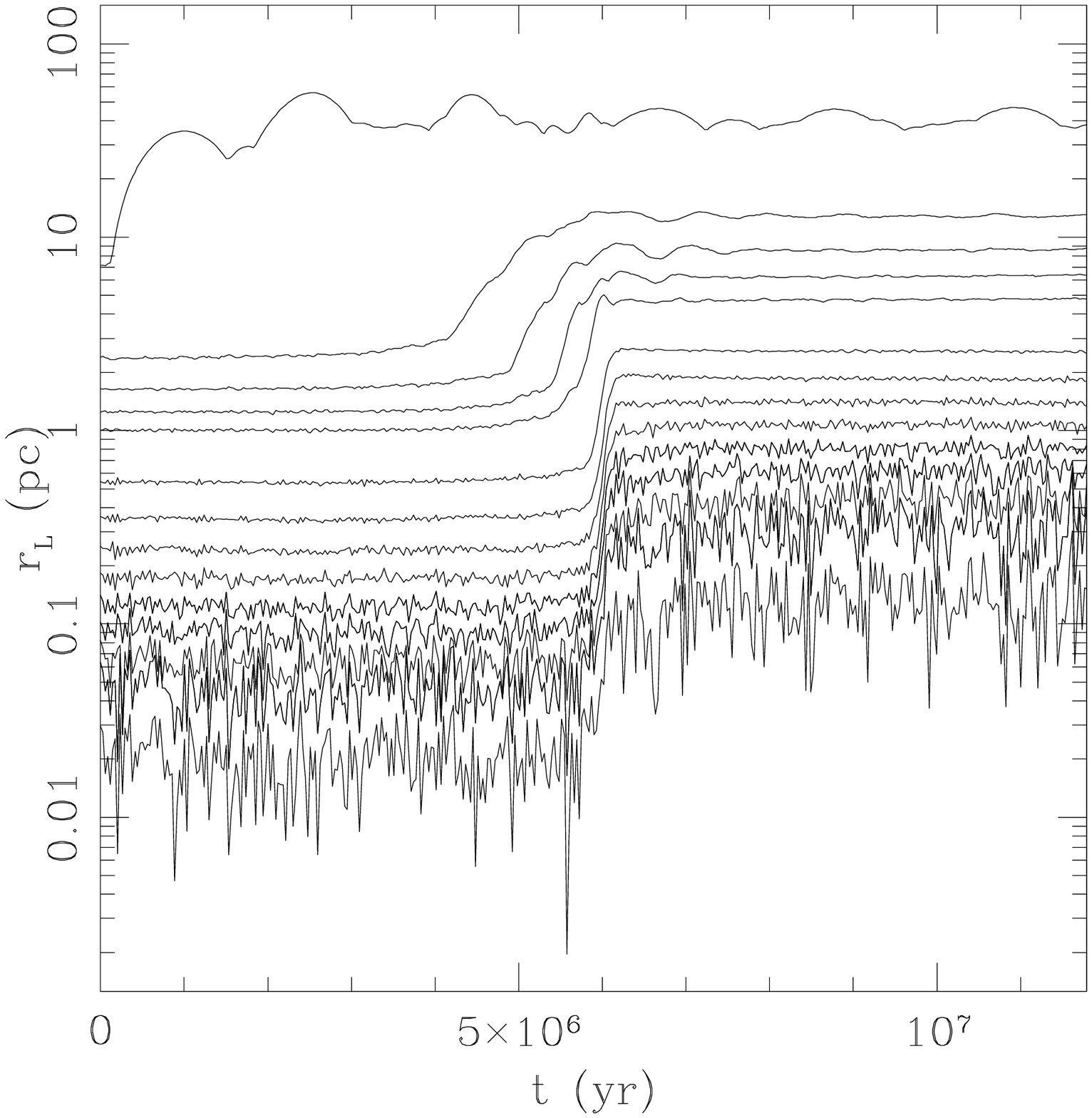}  
\includegraphics[width=0.36\textwidth]{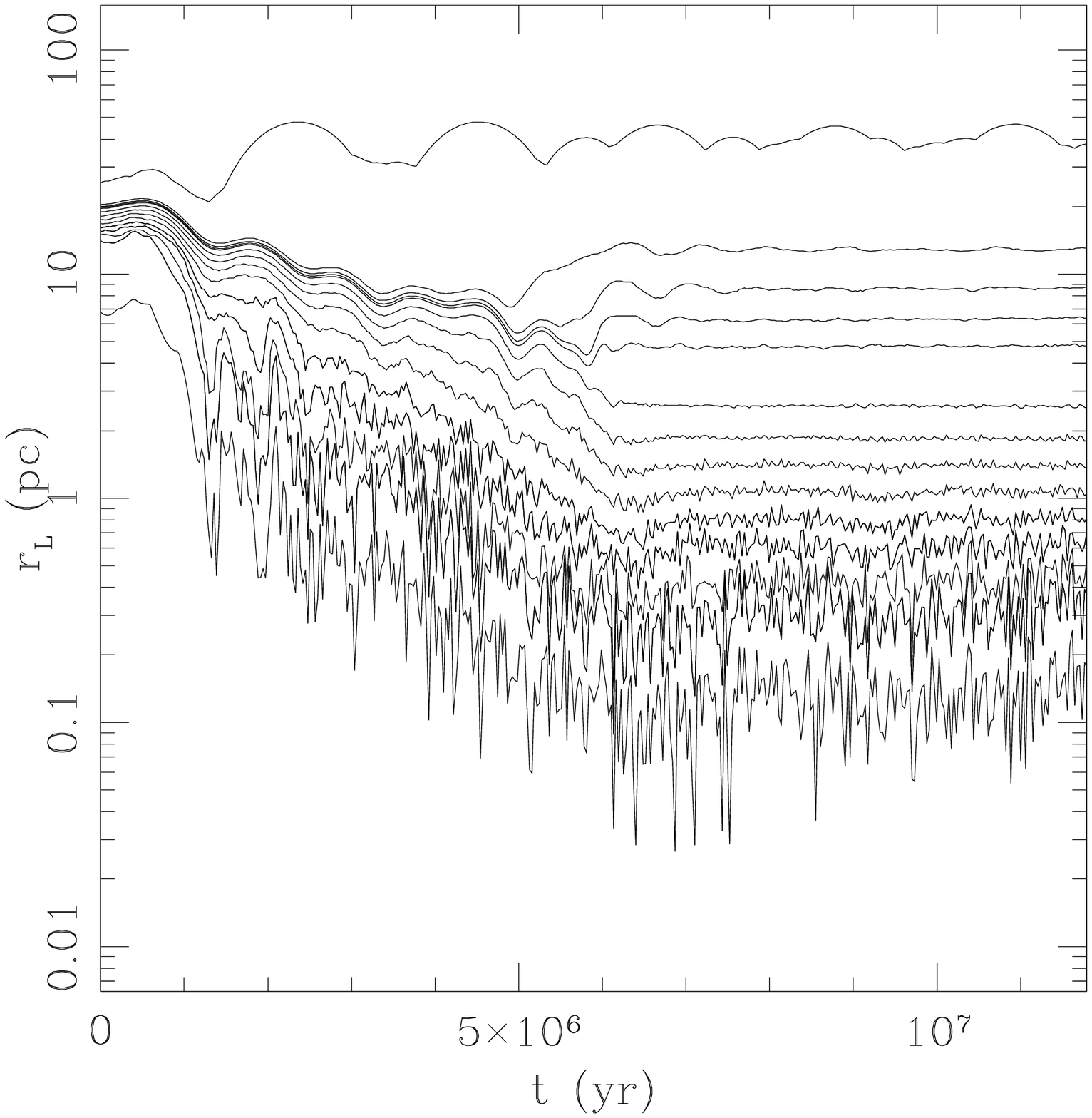}}
\caption{Lagrange radii of stars from the first cluster to arrive at the center of the galaxy. In the upper panel the radii are computed with respect to the center of density of the globular cluster, while in the  lower panel they are computed with respect to the \sbh. The time for each cluster to reach a steady state after its disruption is a few Myr.}\label{fig2}
\end{figure}

\subsection{The Globular Cluster Model}\label{sec:GCmodel}
The globular clusters were initially placed on  circular orbits with orbital radii $r_0=20$~pc.
In order not to favor any particular direction for the inspiral, the orbital
angular momenta were selected in the following way \citep[e.g.,][]{GuaMer09}.
The surface of a sphere can be tessellated by means of 12
regular pentagons, the centers of which form a regular dodecahedron
inscribed in the sphere.
The coordinates of the centers of these pentagons were identified with the
tips of the 12 orbital angular momentum vectors.
In this way, the inclination  and longitude of ascending node  of each initial
orbit were determined.
The choice of circular orbits was motivated by  the well-known effect of orbital
circularization due to dynamical friction \citep{CPV,IL,Hashimoto}.

At a distance of 20 pc from the Galactic center, a globular cluster would already have been subject
to tidal forces from the galaxy and the \sbh, and its total mass and radius
would be less than their original values when the globular cluster was far from the center.
We assumed that the {\it central} properties of the globular clusters were unaffected by tidal forces
during the inspiral to 20 pc, and adopted values characteristic of massive clusters:
central velocity dispersion $\sigma_K=35$~km s$^{-1}$ and
core radius $r_K=0.5$~pc.
If the dimensionless central (King) potential is $W_{0}=8$, the total mass works
out to be $m\approx 4\times10^6~\text{M}_\odot$.
This value of $\sigma_K$ is roughly two times the maximum
value of $\sim 18$km s$^{-1}$ listed in \citet{Harris}'s compilation of Galactic globular clusters properties,
while the core radius is roughly equal to the median value in that compilation.
Our choice of such a large value for $\sigma_K$
is  justified by the fact that only massive clusters, if they are
compact enough, could have arrived in the central regions of the Galaxy in
a reasonable time without being destroyed
by Galactic tidal forces in the process (Miocchi et al. 2006 and \S~\ref{dfts}).

We then needed to generate equilibrium models for globular clusters with these
same central properties, but with total masses and limiting (tidal) radii
consistent with the known tidal forces from the Galaxy model at 20 pc.
This is not a completely straightforward exercise,  since the gravitational
force from the globular cluster acting on a star at the cluster's limiting radius, $r_T$,
depends both on $r_T$ and on the cluster
mass $m_T$ within $r_T$, and $m_T$ is a function of $r_T$.

We proceeded in the following way.
We first assumed $r_T\gg r_K$. In this case, a King-like model satisfies
the following relation between $m_T$ and $r_T$:
\begin{equation}
Gm_{T}\approx \frac12 \sigma_{K}^{2}r_{T} \label{eq:mt}.
\end{equation}
Given this relation, the tidal radius can then be related to the Galactic potential
$\phi$ and density $\rho$ by \citep[e.g.,][]{K62}
\begin{equation}
r_{T}=\frac{1}{\sqrt{2}}\sigma_{K}\left[\frac{3}{r_0}\left(\frac{d\phi}{dr}\right)-4\pi G\rho\right]^{-1/2}\label{eq:rt}.
\end{equation}
Using the galaxy mass distribution of equation (4)
and considering the presence of the \sbh, but ignoring the truncation function,
we find
\begin{equation}
\frac{d\phi}{dr}=\frac{8\pi}{5}G\tilde{\rho} \tilde{r}\left(\frac{r}{\tilde{r}}\right)^{\frac{1}{2}}+\frac{GM_{\bullet}}{r^{2}}~,
\end{equation}
giving a limiting radius of
\begin{equation}
r_{T}=\frac{1}{\sqrt{2}}\sigma_{K}\left[ \frac{4 \pi}{5} G \tilde{\rho} \left( \frac{r_0}{\tilde{r}} \right)^{-1/2}+
\frac{3GM_{\bullet}}{r_0^3} \right]^{-1/2}
\end{equation}
and a tidally-truncated mass from equation~(\ref{eq:mt}).
Adopting a distance  $r_0=20$~pc
we find $r_T\approx 8~$pc and $m_{T}\approx 1.1\times 10^6$~M$_{\odot}$;
in other words, roughly $3/4$ of the globular cluster mass would have been removed in the
process of inspiralling to $20~\mathrm{pc}$.

We then equated this $m_T$ with the mass of a new King model having the same
core properties:
\begin{equation}\label{eq:mTmK}
m_T=m_K\equiv \rho(0) r_K^3 \mu(W_0) \approx
\frac{9}{4\pi G}\sigma_K^2 r_K\mu(W_0).
\end{equation}
Here $\mu(W_0)$ is a function of the dimensionless central potential
that is tabulated by \citet{K66}.
Since all the quantities in equation~(\ref{eq:mTmK}) are known except for
$W_0$, we can solve for this variable, and find $W_0=5.8$.
The three parameters $(W_0,r_K,\sigma_K)$ then uniquely define the King model that was
used  to generate the initial conditions of the globular clusters.

As previously  stated,  the mass of the single particle in the galaxy was $380\text{M}_\odot$.
For the particles in the clusters we choose $200\text{M}_\odot$, approximately one half of that value.
With this choice, the total number of particles in each globular cluster was $5,715$ with   $740$ particles  contained within the cluster core.

\begin{figure}
\centering
\includegraphics[width=0.37\textwidth,angle=0]{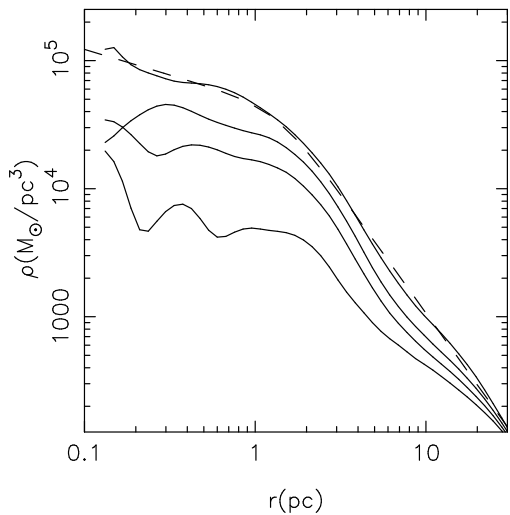}
\caption{Density profile of the NSC after 3, 6, 9 and 12 infall events. 
The central density grows with time. The dashed line
is the fit to the NSC density profile obtained at the end of the entire simulation using the broken power-law model of
equation~(\ref{eq:brokenpl}). }\label{fig:pr_time}
\end{figure}

\section{$N$-body simulations}

Our simulations were performed by using $\phi$GRAPE \citep{Harfst}, 
a direct-summation code
optimized for computer clusters incorporating the GRAPE special-purpose
accelerators \citep{MakinoGRAPE}.
The code implements  a fourth-order Hermite integrator with a predictor-corrector scheme
and hierarchical time stepping.
The accuracy and performance of the code are  set by the time-step parameter $\eta$ and
the  smoothing length $\epsilon$.  In what follows, we set $\eta=0.01$ and $\epsilon=0.02r_K$ ($10^{-2}$~pc in our case),
With these choices, energy conservation was typically
$\lap  0.01\%$ during each merging event.
The simulations were  carried out using  the  32-node  GRAPE cluster at the Rochester Institute of Technology,
and also on computers containing Tesla C2050 graphics processing units at 
Sapienza-Universit\'{a} di Roma.  
In the latter integrations, $\phi$GRAPE was used with {\sc sapporo}, a   {\sc CUDA} library
that  emulates double-precision force calculations on single precision hardware \citep{Gaburov}.

During each infall event, we followed the evolution of the system until the globular cluster had reached the center of the galaxy
and established an approximately steady state.  
This condition was verified by studying the time evolution of the globular 
cluster Lagrange radii, constructed 
both with respect to the center of density of the cluster 
 \citep[as defined by the algorithm in][]{CaHut}, and with respect to the central \sbh.
When the Lagrange radii had reached nearly constant values,
the next globular cluster was introduced.
Figure~\ref{fig2} plots the time evolution of  Lagrange radii for the first infall event.
The figure shows that each merging episode lasts  
approximately $10^7{\rm yr} $ and that the time scale for a globular cluster to reach a steady state following its
disruption   is indeed very short, of the order of Myr.

We  evaluated the $N$-dependence of
our results by  simulating  the first three infalls using the same orbital initial conditions
but with ten times more particles to represent the clusters.
Comparing the density profile of the NSC after the three infalls 
with that obtained in the original integrations did not reveal
any significant differences between the two cases. 

\subsection{Results: density profiles}\label{densityp}
Figure~\ref{fig:pr_time} shows the density profile of the system after the complete merging of 3, 6, 9 and 12 clusters.
We fitted the spatial density  of the final system, within $10$~pc around the \sbh, using the broken power law  model \citep[e.g.,][]{S:92,Z:96}:
\begin{equation}\label{eq:brokenpl}
\rho(r)=\rho_b\left(\frac{r}{r_{b}}\right)^{-\gamma_i}\left[1+\left(\frac{r}{r_{b}}\right)^\alpha\right]^{(\gamma_i-\beta)/\alpha},
\end{equation}
where $\gamma_i$ is the slope of the inner density profile, $\beta$ the external slope and $\alpha$
is a parameter that defines the transition strength between inner and outer power laws.
The best-fit parameters were $\rho_b=4.1\times10^4\text{M}_\odot/\text{pc}^3$, $r_{b}=1.5$pc, $\gamma_i=0.45$,
$\beta=1.90$ and $\alpha=3.73$.
The   model corresponding to this set of parameters is given by the dashed line  in Figure~\ref{fig:pr_time}.

\begin{figure}
\centering
{\includegraphics[width=0.35\textwidth ]{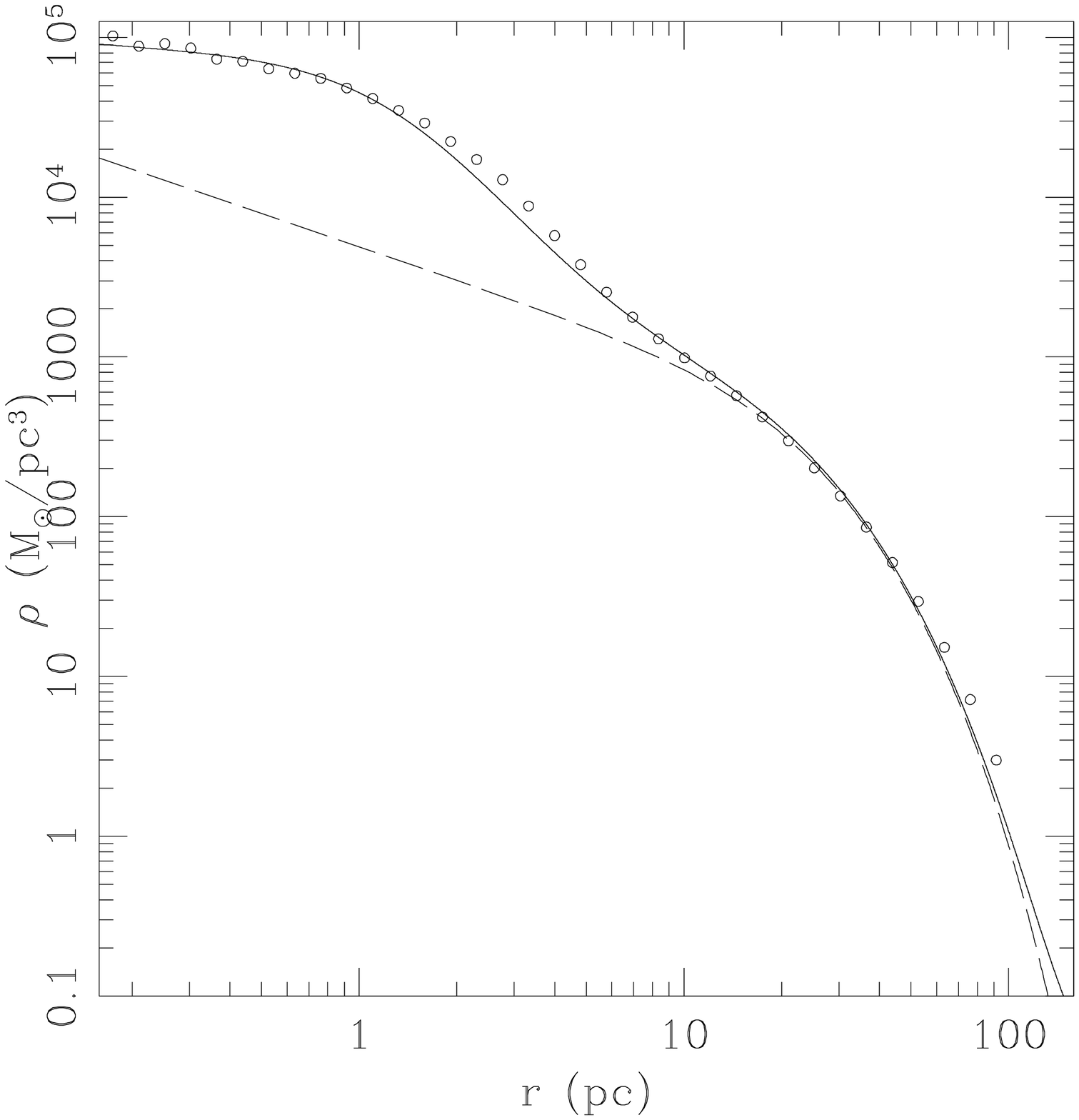}  
\includegraphics[width=0.35\textwidth]{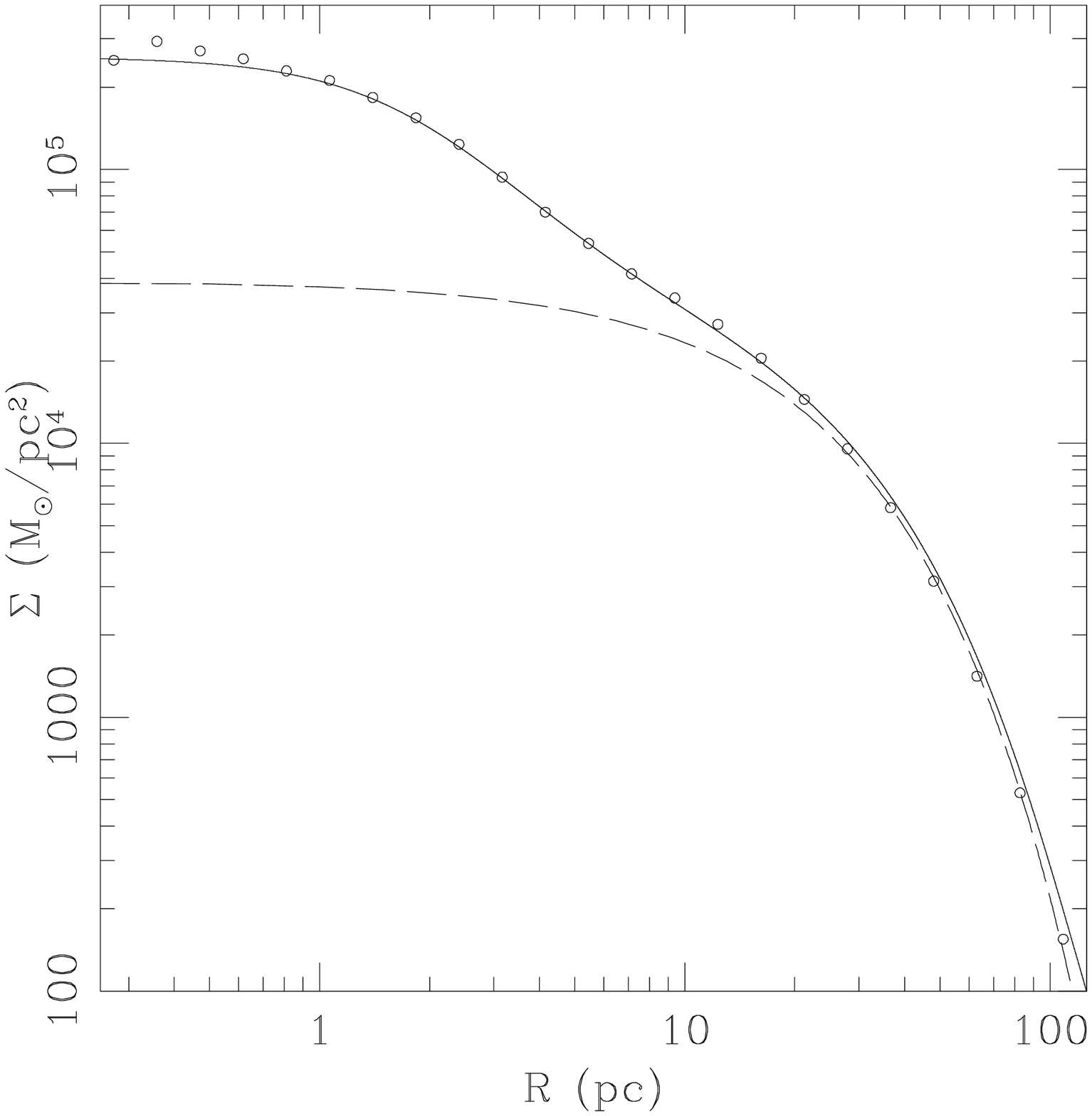} }
\caption{Spatial (upper panel) and projected (lower panel) density profiles at the end of the simulation.
In each panel, the empty circles give the density profile of the $N$-body model,
the solid lines give the best fitting model to the entire system (galaxy+NSC) and
the dashed curves give the fit to the density  profile of the galaxy, see text for explanation.
}\label{fig3}
\end{figure}

\begin{figure}
\centering
\includegraphics[width=0.35\textwidth ]{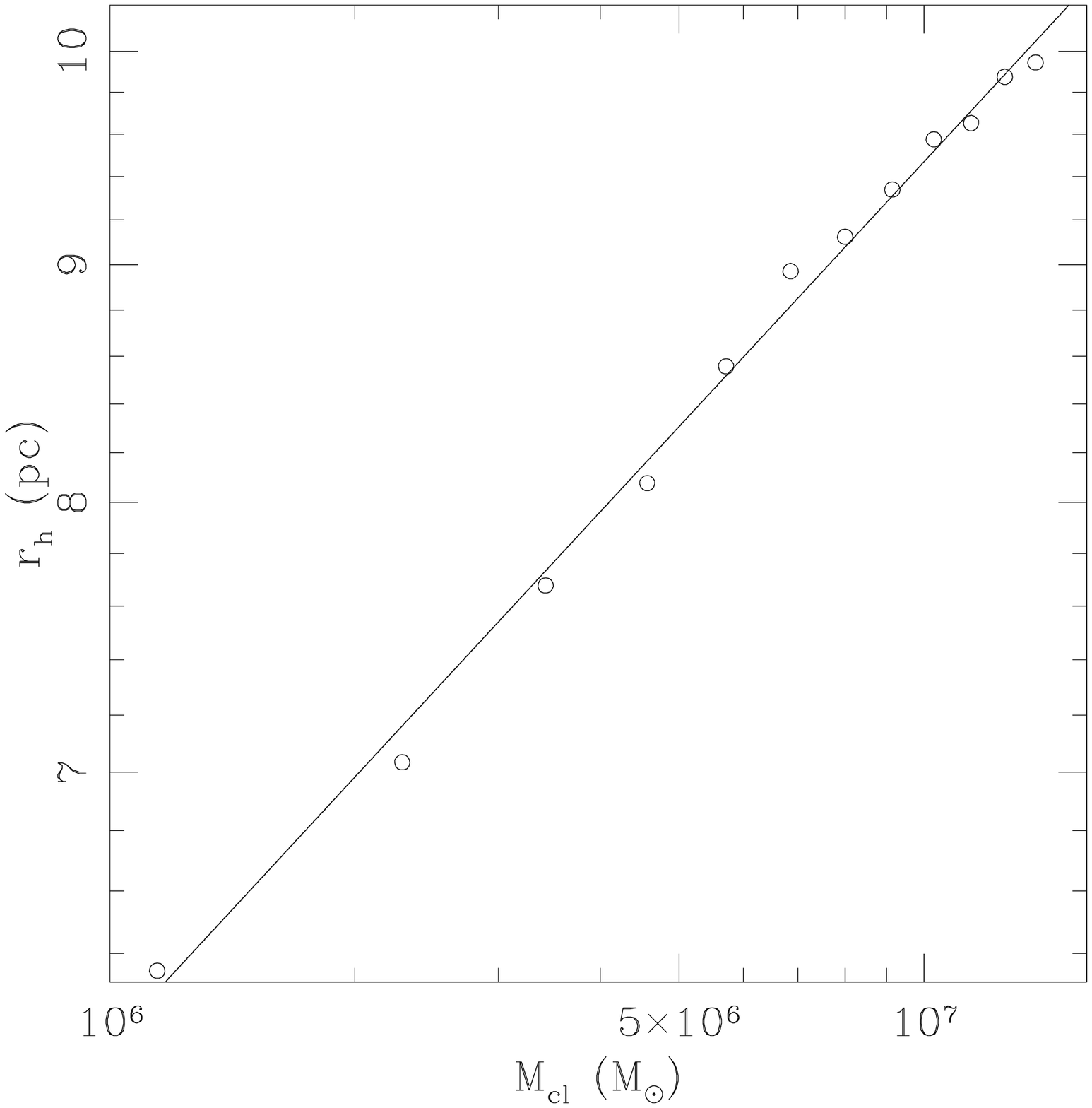}  
\includegraphics[width=0.35\textwidth]{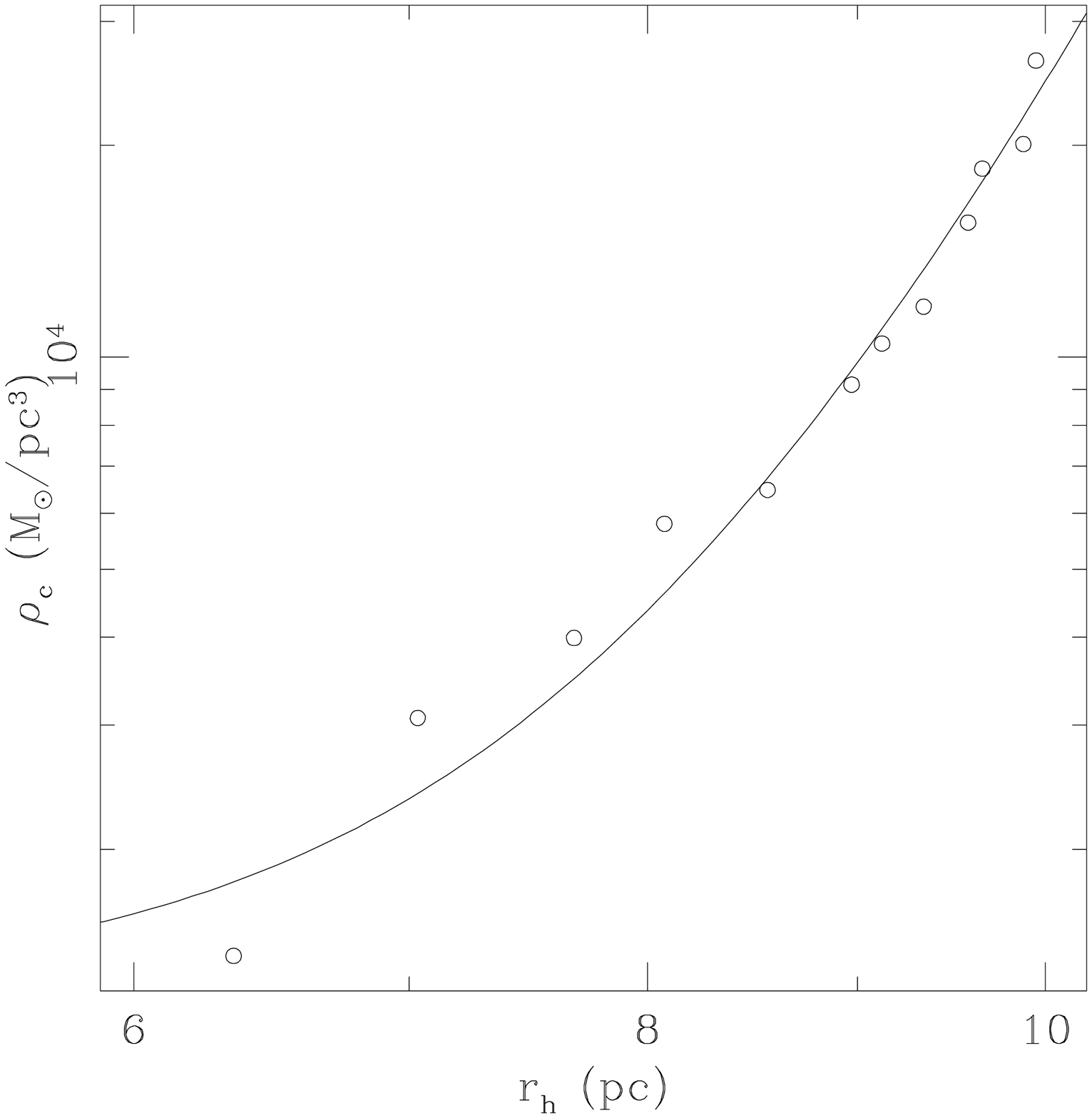}
\caption{Upper panel: empty circles represent the half mass radius of the central NSC as a function of the total mass of the same system.
The solid line represents the scaling relation given in  equation~(\ref{rhM}).
Lower panel: the core density of the cluster versus
its half mass radius (empty circles).  The solid line  shows the  $\rho_c-r_h$ relation given by  equation~(\ref{rhor}).}\label{fig5}
\end{figure}

\begin{figure*}
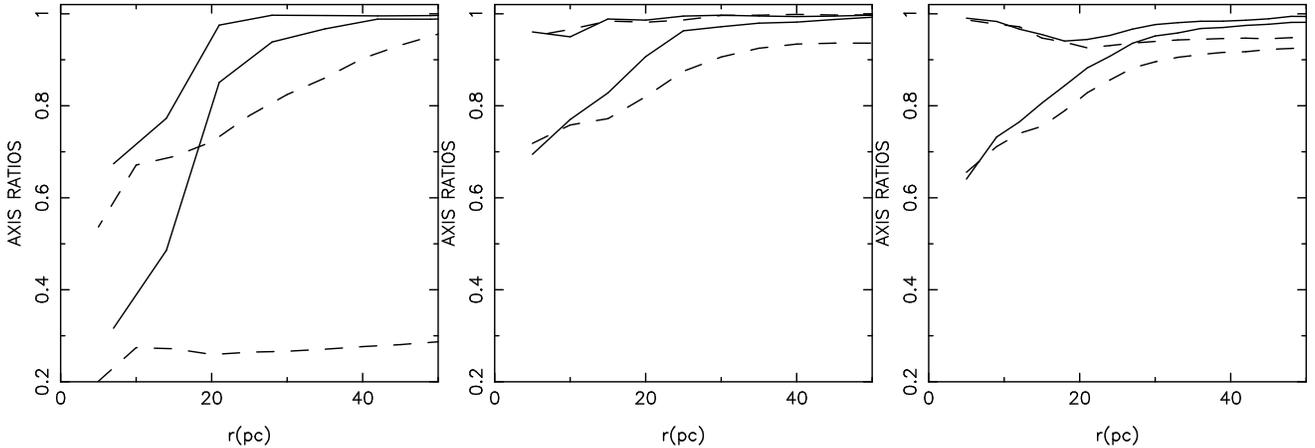

$\begin{array}{lll}
\includegraphics[width=0.32\textwidth]{fig6a.eps} \includegraphics[width=0.32\textwidth]{fig6b.eps}  \includegraphics[width=0.32\textwidth]{fig6c.eps}
\end{array}$
\caption{Axial ratios of the $N$-body system as a function of galactocentric radius computed after $1$ (left panel), $6$ (middle panel)
and $12$ (right panel) infalls.
Solid curves correspond to the entire model (i.e., galaxy plus the NSC), while dashed curves gives the axis ratios of NSC only.
After the first infall the NSC is strongly triaxial in the inner regions, but appears nearly oblate at the end of the simulation.}
\label{shape1}
\end{figure*}

Figure~\ref{fig3} (upper panel) plots the spatial density at the end of the  simulation
over a  wider radial range than in Figure~\ref{fig:pr_time}.  
We fitted the total density as a superposition of two parametric models,
one intended to represent the NSC and the other the galaxy.
For the NSC we adopted the modified Hubble law \citep{R72}:
\begin{equation}\label{eq:mhl}
\rho_{cl}(r)={\rho_{0,cl}}{\left[1+\left( \frac{r}{r_{0,cl}}\right)^2 \right]^{-\frac{3}{2}}},
\end{equation}
with best fitting parameters $\rho_{0,cl}=7.46\times 10^4~\text{M}_\odot/~\text{pc}^3$ and $r_{0,cl}=1.4$~pc.
The galaxy remained well-fit by the initial ``truncated'' power law of equation~(1), when  
$\tilde{\rho}=9.91\times10^2~\text{M}_\odot/~\text{pc}^3$, $\tilde{r}=10$~pc, $\gamma=0.69$,   and $r_{\rm cut}=16.3$~pc.

The lower panel of Figure~\ref{fig3}  
shows the projected density of the $N$-body model at the end of the simulation.
We again fit this profile with a two-component model.
For the NSC we used 
\begin{equation}\label{eq:mcm}
\Sigma_{cl}(R)={\Sigma_{0,cl}}{\left[1+\left(\frac{R}{R_{0,cl}}\right)^2\right]^{-\zeta}}~,
\end{equation}
while the projected density profile of the galaxy was represented by a  S\'ersic law:
\begin{equation}\label{eq:sersic}
\Sigma_{gx}(R)=\Sigma_{0,gx} \text{exp}{\left[ -b\left( \frac{R}{R_{0,gx}} \right)^\frac{1}{n} +b\right]}~,
\end{equation}
with
\begin{equation}
b=2n - \frac{1}{3} + \frac{0.009876}{n}.
\end{equation}
The best-fit parameters were $\Sigma_{0,cl}=2.18\times10^5~\text{M}_\odot/\text{pc}^2$,
$R_{0,cl}=1.99$~pc and $\zeta=1.03$ for the NSC;  $\Sigma_{0,gx}=7.31\times10^3~\text{M}_\odot/~\text{pc}^2$,
$b=1.68$, $R_{0,gx}=32.3$~pc and $n=1.003$ for the bulge.

Remarkably,  our simulations result in a final density profile having nearly the same 
power-law index beyond $\sim 0.5$~pc as observed
\citep[$\Sigma(r) \sim r^{-1}$;][]{Becklin,Haller}.
In addition,  the central region ($r<r_b$) of our model exhibits a shallow
density profile (or core) near the \sbh, also in agreement with observations \citep{Buch}.  
The core radius in our model ($\sim 2$~pc) is somewhat larger than the observed
core, of radius $\sim 0.5~$pc.
In \S \ref{sec:evol} we show that two-body relaxation would cause such a core
to shrink over Gyr time scales, as the density evolves toward, but does not fully
reach, a collisional steady state.

In the upper panel of Figure \ref{fig5}  the half-mass radius ($r_h$)  of the NSC component
is plotted as a function  of the NSC   mass ($M_{cl}$) at the end of each infall.
At any time, the NSC mass is given by the sum of the accumulated globular cluster masses.
A good fit to the data is obtained by 
\begin{equation}
r_h=0.45\left(\frac{M_{cl}}{M_\odot}\right)^{0.19}~\text{pc}~, \label{rhM}
\end{equation}
represented by the solid line.
The dependence of $r_h$ on $M_{cl}$ is weak, due to the fact that the size of the
NSC is determined essentially by the fixed tidal field from the \sbh.

Assuming for the growing NSC the  density law of equation (\ref{eq:mhl}),
the core density can be defined as  $\rho_c=\rho_{0,cl}/2^{\frac{3}{2}}$ and the values
of $\rho_c$, obtained  after the end of each infall, can be plotted as a function
of the half mass radius of the same system  (bottom panel of Figure~\ref{fig5}). These data are well fit by
\begin{equation}
\rho_c=\left[1.2\times 10^3 + 1.1 \exp\left({\frac{r_h}{1~\text{pc}}}\right) \right]
\frac{\text{M}_\odot}{\text{pc}^3}~,\label{rhor}
\end{equation}
shown as solid curve in the figure.

We also carried out a separate simulation in which all 12 clusters were placed at the same time on their initial orbits.
Such initial conditions are hardly realistic; 
we mention the outcome briefly here since it supports
the robustness of the results obtained from the more realistic initial conditions.
The ``contemporaneous'' infall model
produced a very similar final density profile, with a core of radius $\sim 1$ pc and 
a $\rho\sim r^{-2}$ falloff at larger radii.

\subsection{Results: morphology of the NSC}\label{ar}
Observationally constraining the morphology and kinematics  of galactic nuclei
is a fundamental step toward  understanding their origin.
Unfortunately, as a consequence of the  strong interstellar extinction in the plane
of the Milky Way,
our knowledge of the size and morphology of the Galactic NSC are limited.
Kinematic modeling of proper-motion data derived from the dominant  population of old, mostly giant stars reveals a nearly spherical system of low central concentration
exhibiting  slow, approximately solid-body rotation, of amplitude $\sim 1.4 {\rm km~s^{-1}/arcsec}$
\citep{trippe,S09}.

Aspherical NSCs are commonly observed in external galaxies.
In a sample
of 9 edge-on  nucleated late-type galaxies, \citet{SETH06} reported  that
three of these galaxies   (IC 5052, NGC 4206, and NGC 4244) have  NSCs with  significantly flattened isophotes and evidence for multiple structural components.
In addition, one of these galaxies  (NGC 4206)  showed possible indication of AGN activity,
suggesting the presence of  a \sbh\  within the core of the central cluster.
The  NSC of the face-on galaxy M33, in which a \sbh\ is not detected
\citep{MerrittM33,GebhardtM33},
is also known  to be elongated  along an axis parallel to the
major axis of the galaxy \citep{LAUER,Matthews}.

We quantified the model shape in our simulation   by
constructing isodensity contours  and also by the moment-of-inertia tensor~\citep[e.g.,][]{Katz:1991,PM:04, ACM09},
as described in what follows:
the  symmetry axes are calculated as
\begin{equation} \label{rapp_ax}
\tau_1=\sqrt{I_{11}/I_{max}}~,~\tau_2=\sqrt{I_{22}/I_{max}}~,~\tau_3=\sqrt{I_{33}/I_{max}}~,
\end{equation}
where $I_{ii}$ are the principal moments of the inertia tensor
and $I_{\mathrm {max}}={\rm max}\{I_{11},I_{22},I_{33}\}$;
particles are then enclosed within the ellipsoid $x^2/{\tau_1}^2+y^2/{\tau_2}^2+z^2/{\tau_3}^2=r^2$.
These previous two steps were iterated  until the values of the axial ratios   had a percentage change of less than $10^{-3}$.
Finally, we define $a > b > c$ letting $c/a=\min\{ \tau_1, \tau_2, \tau_3 \}$ and  $b/a$ the intermediate value.
We also define the triaxiality via the parameter $T\equiv \left(a^2-b^2\right)/\left(a^2-c^2\right)$.
Oblate and prolate galaxies have $T=0$ and $1$, respectively.
The value $T=0.5$ corresponds to the \lq maximally triaxiality\rq ~case.

The results are summarized in Figure~\ref{shape1} which displays  the axial ratios of the NSC
as a function of radius and at different times.  
The model morphology evolves from an initially strong triaxiality   (after the first infall)
into a more quasi-axisymmetric, oblate shape.
In particular, the morphological structure of the final product (right panel)
is very similar  to that after  the  6th infall event (middle panel).
This shows that the NSC is  transformed into a  nearly oblate system ($T \lesssim 0.2$ at $r<20~$pc)  after few infalls ($\sim 4$), but
its shape  remains essentially unchanged from that point on.
In the outer regions ($\gtrsim 20~$pc), the system remained  instead  nearly spherical for the entire course of
the simulation.

\subsection{Results: kinematics}\label{sub:kin}
\begin{figure*}
$\begin{array}{rr}
~~~~~~~~~~~~~~~~~~~~~~~~~~\includegraphics[width=0.35\textwidth ]{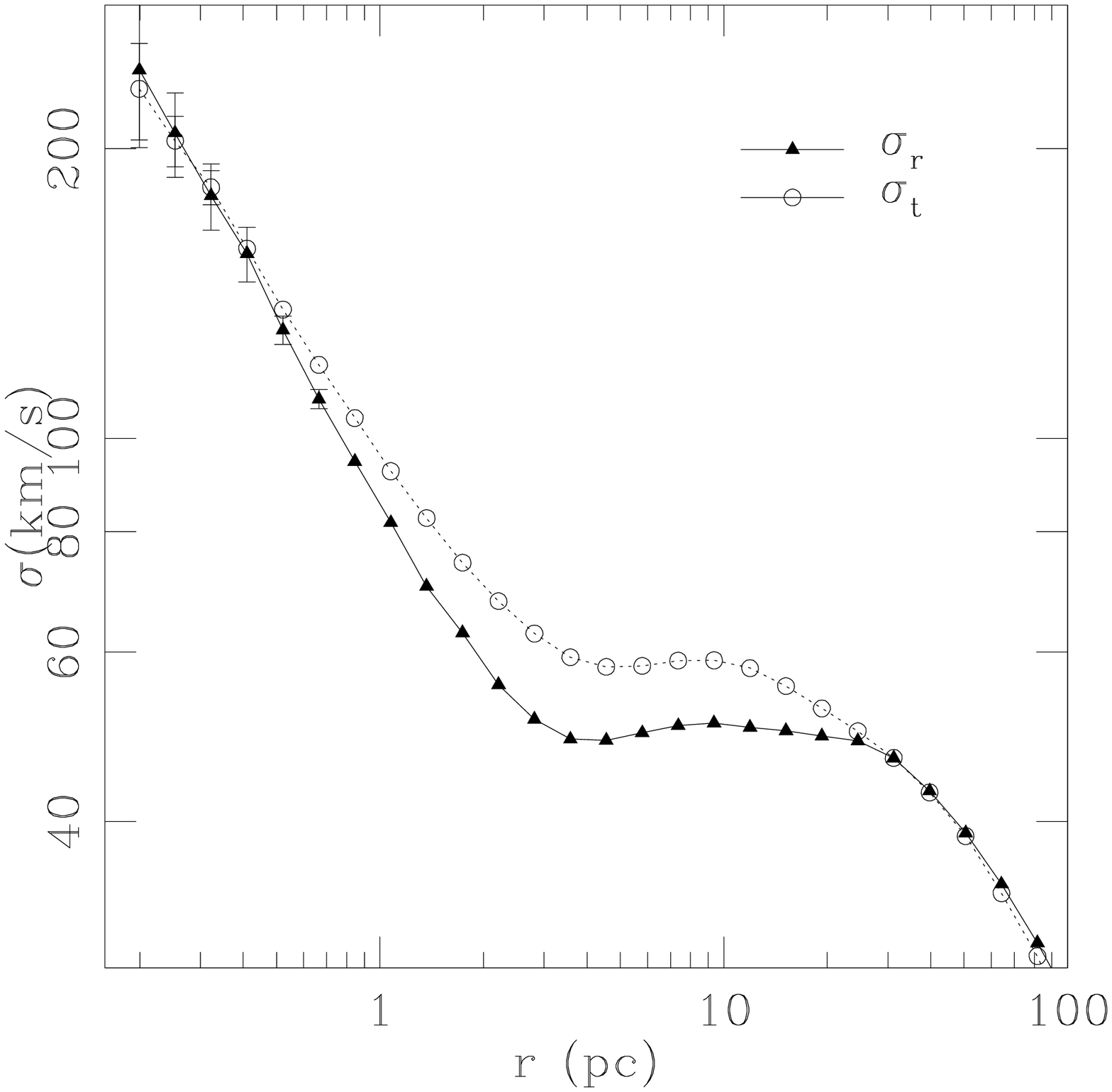}  \includegraphics[width=0.35\textwidth]{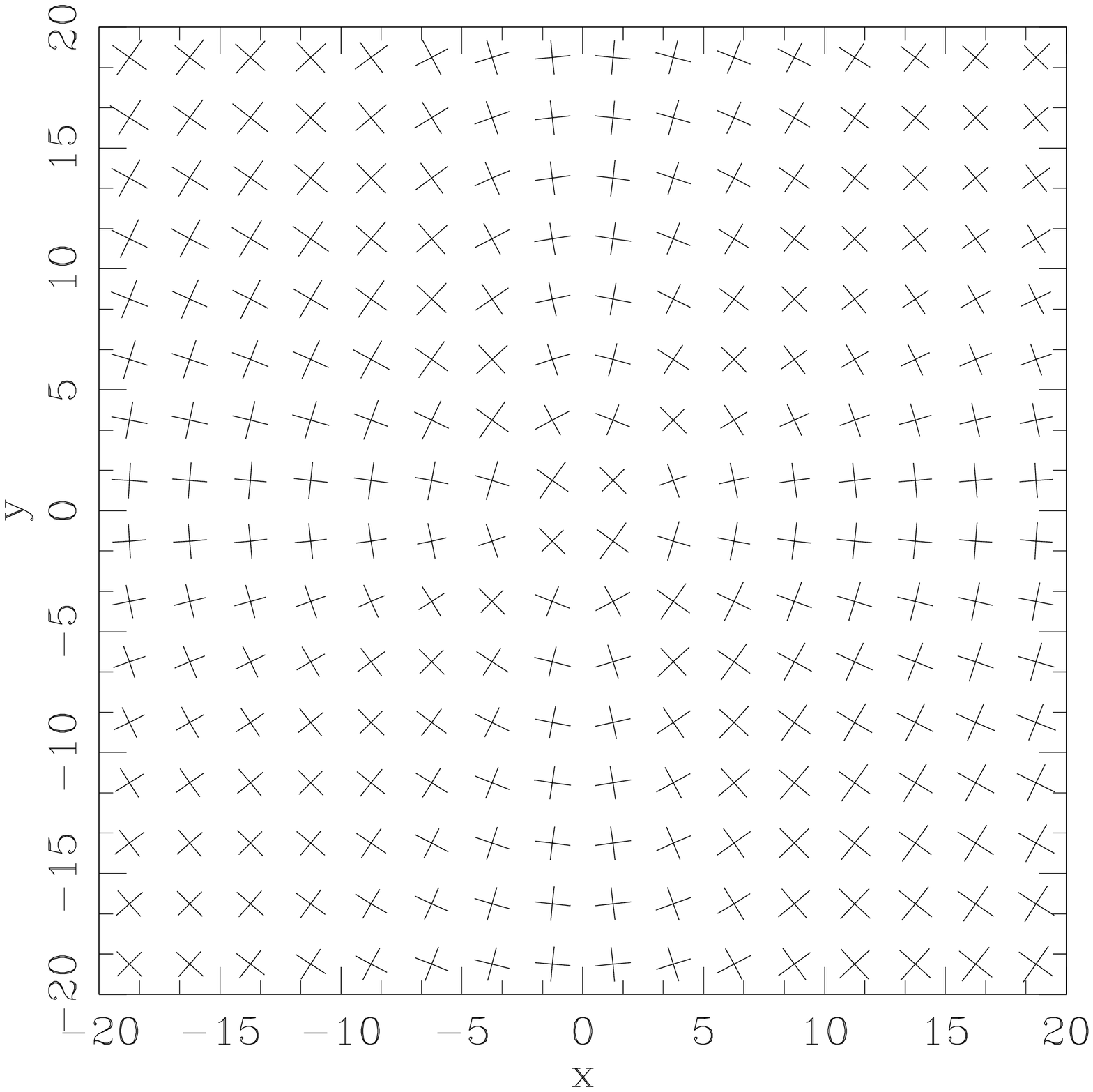} \\
~~~~~~~~~~~~~~~~~~~~~~~~~~\includegraphics[width=0.35\textwidth ]{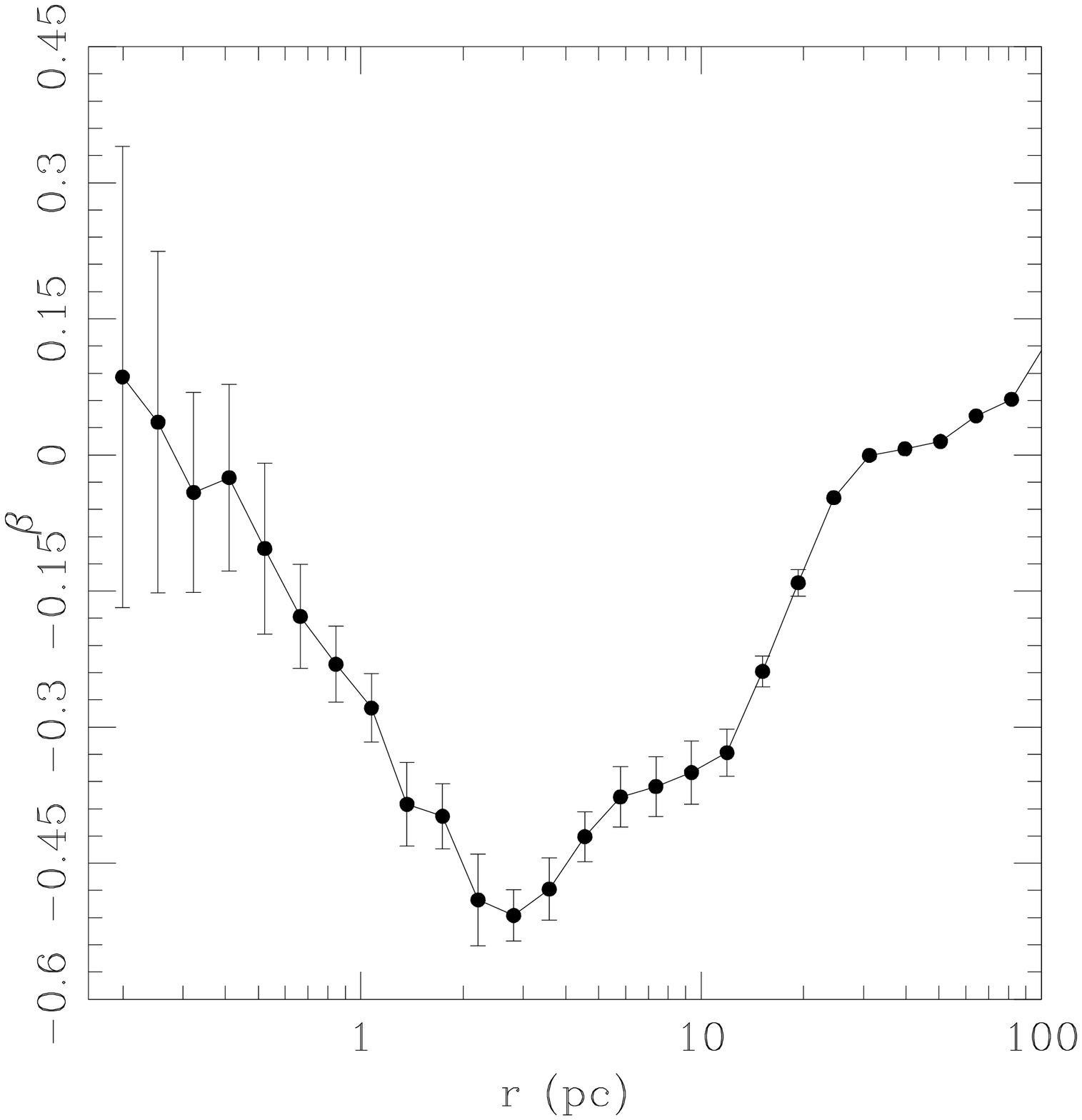}\includegraphics[width=0.35\textwidth]{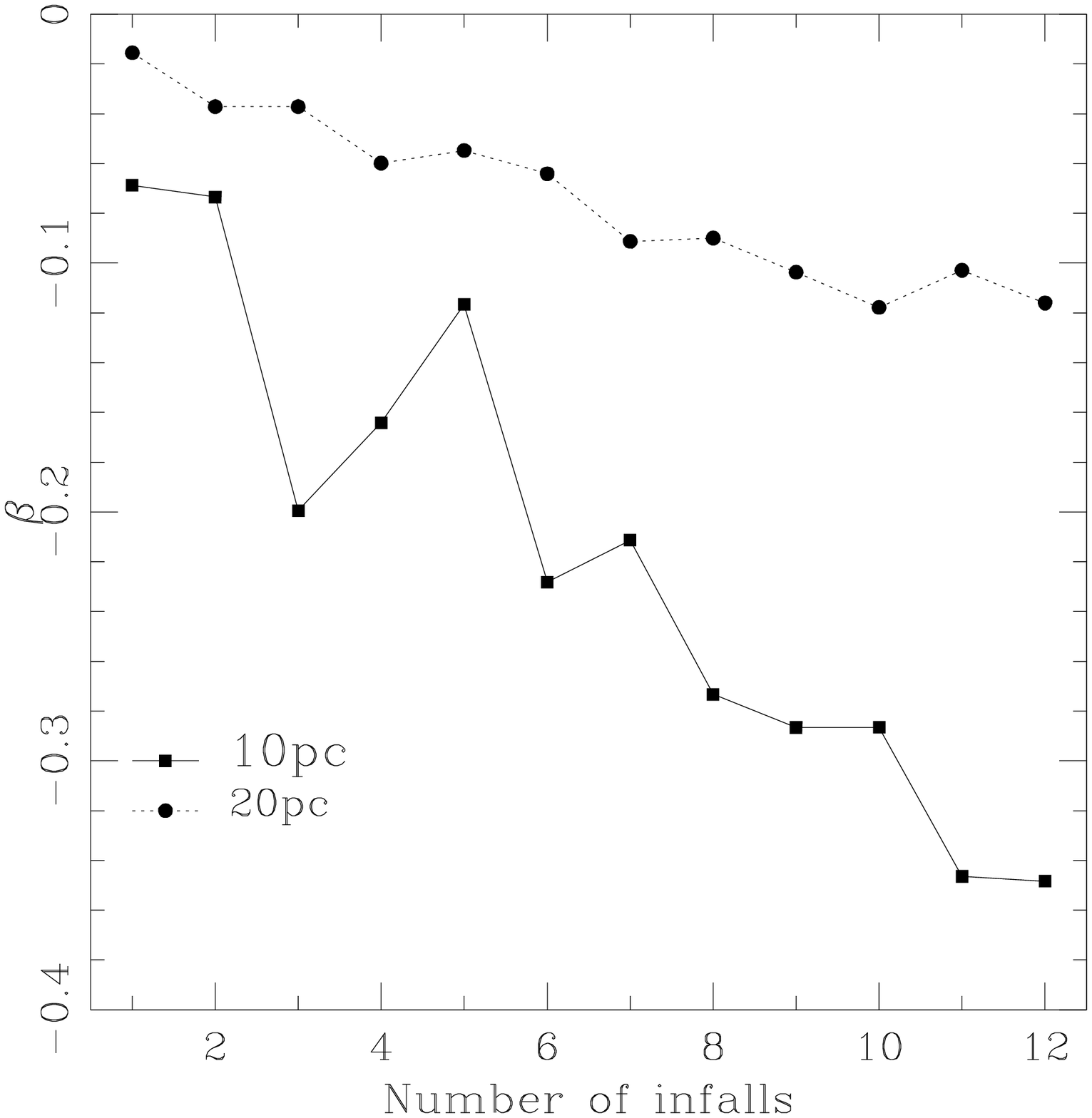}
\end{array}$
\caption{{\it Upper left:} radial ($\sigma_r$) and tangential ($\sigma_t$) 
velocity dispersions plotted as functions of the distance from the \sbh.
{\it Upper right:} map of the principal axes of the 2D velocity ellipses derived
from the $(x,y)$ velocity components.
{\it Lower left:} Anisotropy parameter $\beta$.
{\it Lower right:} The anisotropy
parameter evaluated at $10$ and $20$~pc versus  time.
All  plots were derived form the model
at the end of the 12th inspiral event.}\label{fig7}
\end{figure*}

Figure \ref{fig7} illustrates the kinematics of the final model.
The upper-left panel of the figure shows the one-dimensional velocity dispersions
along, and perpendicular to, the radius vector, defined with respect to the \sbh.
For $r<0.3$~pc the system is quite isotropic, but it becomes tangentially anisotropic
for $0.3\mathrm{pc}<r<20$~pc.
At larger radii the system is again roughly isotropic.
The upper right panel of Fig. \ref{fig7}  shows the amplitude and orientation of the
major axes of the 2d velocity ellipsoid, derived from the $(x,y)$ velocities.
The length of the plotted axes  is proportional to the velocity dispersion.
The tangential anisotropy is apparent.

The lower panels of Figure \ref{fig7} show the anisotropy parameter
\begin{equation}
\beta=1-\frac{\sigma_t^2}{2\sigma_r^2}~,
\end{equation}
as a function of distance from the \sbh,
and as a function of time at two different distances, $10$~pc and $20$~pc.
In the radial range $0.3- 40$~pc, $\beta$ is negative and the system is tangentially anisotropic.
The anisotropy grows with the number of infall events.

\section{Collisional evolution of the Nuclear Star Cluster}\label{sec:evol}
The simulations of globular cluster inspiral described above
took place in a short enough span of time that two-body relaxation effects could be ignored.
The local relaxation time can be defined as \citep{Spitzer}
\begin{eqnarray}\label{tre}
t_\mathrm{r}=\frac{0.33 \sigma^3}{ G^2 \rho m_\star \ln\Lambda}
\end{eqnarray}
where $m_\star$ is the stellar mass, $\rho$ is the mass density and $\sigma$
is the one-dimensional velocity dispersion.
Near the influence radius of a \sbh, the Coulomb logarithm can be approximated as
$ \ln\Lambda= {\rm ln}(r_{\rm infl}\sigma^2/2G m_*)$.
The relaxation time at the
influence radius of Sgr A$^*$,
$r_\mathrm{infl}=2-3\:\text{pc}$,
is $t_\mathrm{r,infl} \sim20-30\:\text{Gyr}$, assuming a stellar mass
of $1M_\odot$ \citep{Mer10}.
This is roughly $2\times 10^5$ times the period of a circular orbit
at $r_\mathrm{infl}$.
In our $N$-body simulations, the relaxation time is shorter (compared
with the crossing time) by a factor of approximately 200, the mass of
a single cluster particle in solar masses; in other words, it is roughly
$10^3$ times the crossing time at $r_\mathrm{infl}$.

In the absence of large-scale changes to the gravitational potential, an $N$-body model
like ours continues to evolve due to gravitational encounters.
The evolution that occurs should mimic the evolution that would take place
in the real system, of much larger $N$, if the unit of time is taken to be the
relaxation time~\citep[e.g.,][]{AAR98}.

\begin{figure}
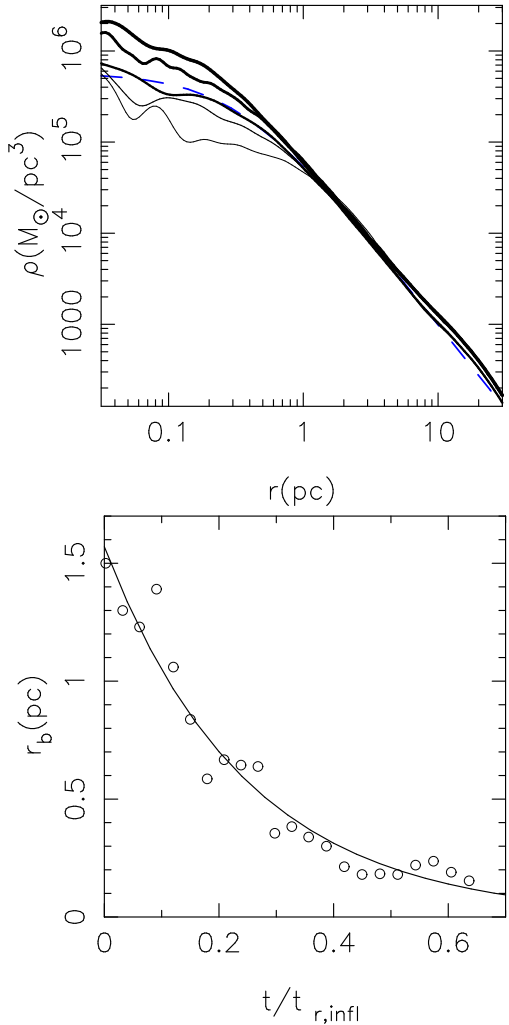

$\begin{array}{rr}
~~~~~\includegraphics[scale=.6]{fig8a.eps} \\ \includegraphics[scale=.6]{fig8b.eps}
\end{array}$
\caption{Post-infall evolution of the $N$-body model.
Upper panel shows the density profile at four times:
$t=(0.12, 0.24, 0.36, 0.48, 0.6)$ in units of the relaxation time at the
influence radius; line thickness increases with time.
The dashed (blue) curve shows a fit of the density to
the broken power-law model of equation (\ref{eq:brokenpl})
at $t\approx 0.36~t_\mathrm{r,infl}$, i.e., $\sim 10~$Gyr when scaled to the Milky Way.
Lower panel plots the break (core) radius as a function of time.
The solid line is the best-fit exponential, equation~(\ref{eq:rbVSt}).} \label{fig:ev_time}
\end{figure}

In the Milky Way, the relaxation time is short enough that significant
evolution of the stellar distribution near the \sbh\ would take place over the age of the
galaxy.
The distribution of late-type stars in the Milky Way NSC exhibits a nearly flat core
of radius $\sim0.5\:{\rm pc}$ \citep{Buch}.
In a time of $10$ Gyr, such a core would shrink,
as the stellar density evolved toward the Bahcall-Wolf (1976)
$\rho\sim r^{-7/4}$ form inside $\sim 0.2 r_\mathrm{infl}$.
Since the density profile beyond the core is observed to
have roughly this slope \citep{Oh}, such evolution would tend to preserve
the outer slope while gradually reducing the size of the core.
A core of initial radius $1-2$ pc is expected to reach a size of
$\sim 0.5$ pc, the size of the observed core, after $\sim 10$ Gyr \citep{Mer10}.
These arguments motivated us to continue the integration of our $N$-body models
after the final inspiral event.
Figure \ref{fig:ev_time} shows the density profile
of the NSC at different times during its post-merger evolution.
At the end of this integration,
i.e. after $\sim 0.6\, t_\mathrm{r,infl}$,
the distribution shows an inner core of size $\sim 0.2$ pc,
substantially reduced from its initial value of $\sim 1.5$ pc.
The bottom panel of Figure \ref{fig:ev_time} plots the evolution of the break
radius, $r_{b}$, of the best fitting broken power law profile as a function of time.
The value of the break radius can be used  as an approximate estimate  of the model core radius.
The time dependence of the core radius is well described by an exponential:
\begin{equation}\label{eq:rbVSt}
r_{b}(t)=1.57 \mathrm{pc}\exp\left[-t/(0.25 t_\mathrm{r,infl})\right].
\end{equation}
As expected, the slope of the density profile outside the core remains nearly
unchanged during this evolution, $\rho \simeq r^{-1.8}$.

A core radius of $\sim 0.5$ pc is reached after a time of
$\sim 0.25 t_\mathrm{r,infl}$.
Scaled to the Milky Way, this time would be $5-8$ Gyr.

Of course, in the real galaxy, it is likely that cluster inspiral would
occur more or less continuously over the lifetime of the galaxy.
Our separation of the evolution into an inspiral phase, followed by
a relaxation phase, is artificial in this sense.
Nevertheless it is reasonable to draw the conclusion that the size
of the core resulting from the combined effects of cluster inspiral and
relaxation would be somewhat smaller than the $\sim 1.5$ pc that we
found above, and therefore, closer to the observed core size of $\sim 0.5$ pc.

\begin{figure}
~~~~~~~~\includegraphics[width=.35\textwidth]{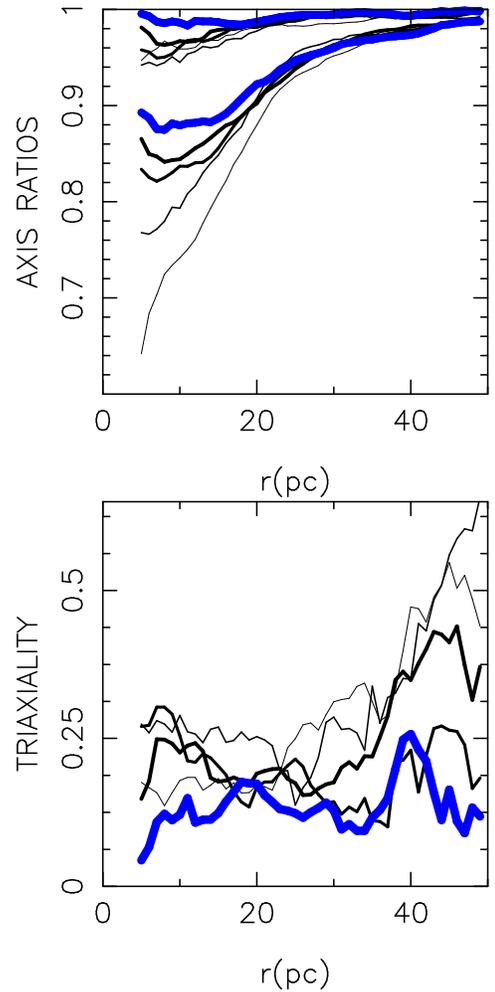}
\caption{Evolution of the model axis ratios (upper panel) and
triaxiality parameter (lower panel), as functions of radius, in the
post-merger phase.
Times are the same as in Figure~\ref{fig:ev_time};
line thickness increases with time. Thickest blue lines correspond to the the final model.}\label{figs}
\end{figure}

Figure~\ref{figs} shows the morphological evolution of the NSC during the relaxation
phase: the radial dependence of the axis ratios (upper panel) and the
triaxiality parameter (lower panel).
There is essentially no evolution in the intermediate axis ratio.
However, in the innermost regions of the model, the shortest axis length significantly
increases with time. Two-body relaxation results in an evolution toward quasi-spherical symmetry, but at the end of the simulation
the model has not yet reached this final state, still exhibiting some non-negligible triaxiality.
The final model is nearly oblate with $0.3\lesssim T \lesssim0.1$.
We note that such deviations from spherical symmetry, although mild, might be large
enough to substantially enhance the rates of stellar capture and disruption by the \sbh\
with respect to the same rate computed in collisionally resupplied loss cone theories \citep{Mer12}.

\begin{figure}
~~~~~~~~~~\includegraphics[width=.37\textwidth]{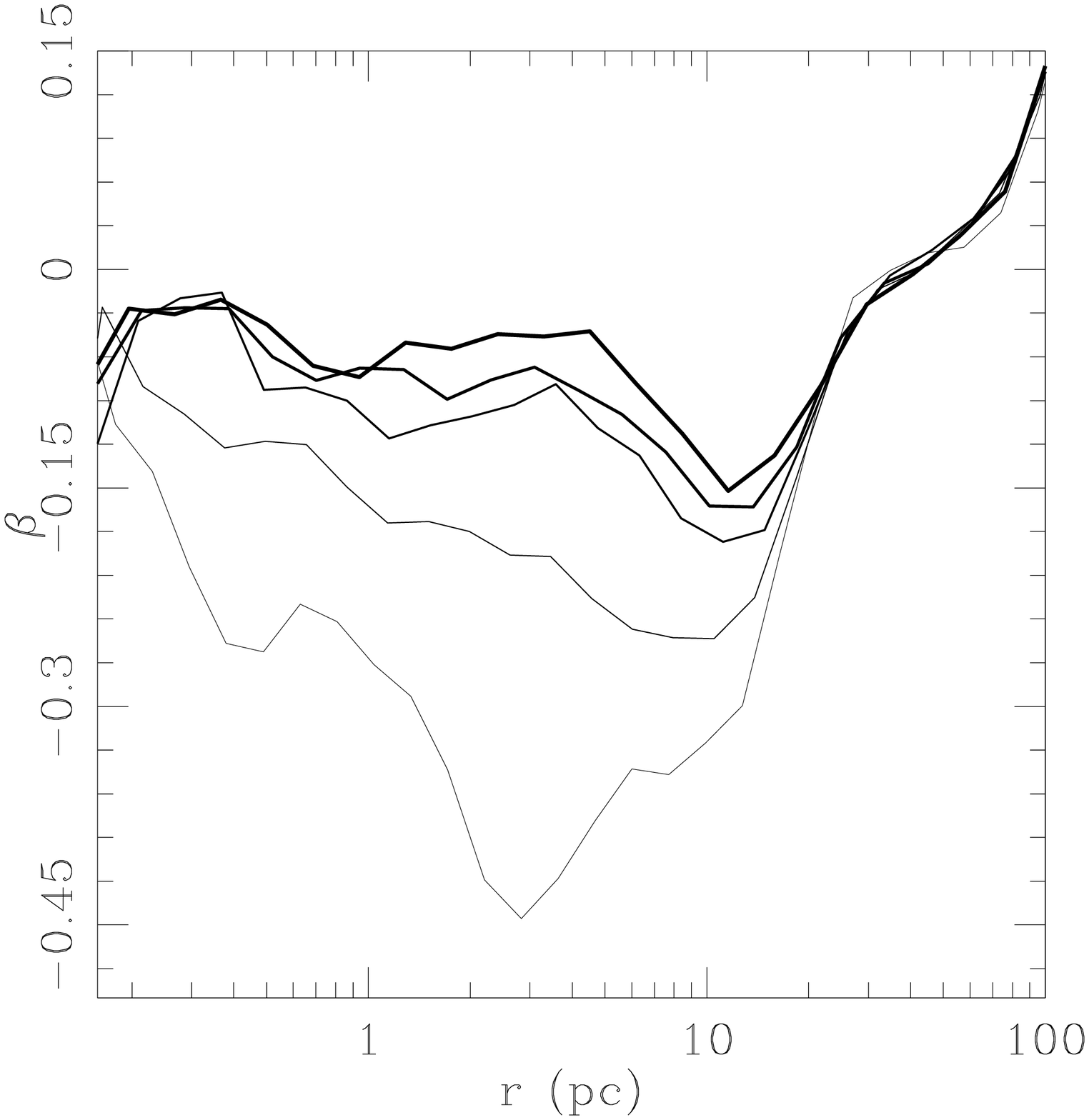}
\caption{Evolution of the anisotropy parameter $\beta$ during the post-merger
phase.
Times shown are the same as in Figure~\ref{fig:ev_time}; line thickness increases with time.
}\label{fig:beta}
\end{figure}

Figure~\ref{fig:beta} illustrates the evolution of the velocity anisotropy profile,
$\beta(r)$, during the post-merger phase.
Relaxation tends to drive the velocity distribution toward isotropy,
causing $\beta$ to increase toward zero~\citep[this dynamical trend is similar to that described in][]{Capuc}.
After $0.4t_\mathrm{r,infl}$, i.e., $\sim10~$Gyr,
there remains only a small bias toward tangential motions, $\beta \sim -0.1$,
$r\lap 10$ pc.
The final anisotropy profile is consistent with measurements of the
Galactic center \citep{S09,Mer10}.
In the radial range $1\arcsec-10\arcsec $, the late-type stars
are observed to have a mean projected anisotropy of
$1-\langle\sigma_T^2 / \sigma_R^2 \rangle=-0.124_{-1.05}^{0.098}$
with $\sigma_R$ and $\sigma_T$ the radial and tangential velocity dispersions in the plane of the sky.

\section{discussion}

\subsection{Cluster inspiral times}\label{dfts}

For the model described here to be viable, globular clusters must spiral into the central regions of their parent galaxy in a time shorter than $\sim 10^{10}~$yr.
A rough estimate of the time for a globular cluster of mass $m$ at initial radius $r_0$ to inspiral into the center of a galaxy as a result of dynamical friction is \citep{M04}:
\begin{equation}
\Delta t= \frac{C(\gamma)}{\rm ln \Lambda} \frac{r_a^3}{m}\left( \frac{\rho}{G} \right)^{1/2} \left( \frac{r_0}{r_a} \right)^{(6-\gamma)/2}~.\label{dftse}
\end{equation}
This expression  assumes circular orbits, a total mass density of the galaxy that satisfies $\rho(r)=\rho_a(r/r_a)^{- \gamma}$,
and a frictional force that is due entirely to stars with velocities less than the orbital velocity of the globular cluster,
and Coulomb logarithm ${\rm ln} \Lambda=6.5$ \citep{AM:2011}.
The coefficient $C$ in equation~(\ref{dfts}) is a weak function of $\gamma$, with $C=(3.9, 3.6, 4.2)$  for $\gamma=(1, 1.5, 2)$.
Writing  $\tilde{r}_a\equiv r_a/(1 {\rm kpc})$, $\tilde{\rho}_a \equiv \rho_a/(1 M_{\odot}pc^{-3})$, $\tilde{m}\equiv m/ 10^6 M_{\odot}$
and $\tilde{r_0}\equiv r_0/r_a$, the time for a globular cluster to spiral in is:
\begin{equation}
\Delta t  \approx 9 \times 10^9 {\rm yr}~~ \tilde{r}_a^3 \tilde{\rho}_a^{1/2} \tilde{m}^{-1} \tilde{r}_0^{(6-\gamma)/2}~.
\end{equation}
Although the initial distribution of globular clusters is not known, a reasonable assumption
is that it follows the distribution of the total baryonic mass predicted by the standard cosmological model.

Let $\Delta t_{h}$ be the time for clusters initially within $r_{h}$, the half mass radius of the galactic bulge,
to spiral to the center. Within $\Delta t_{h}$, the forming nucleus has a luminosity comparable
to the luminosities of the surviving clusters. An estimate of $r_{h}$ is $r_{e}$,
the observed projected half-light (effective) radius; the stellar density at $r=r_{e}$
is $\tilde{\rho}_e\approx 14.1 \tilde{r}_e^{-2.519}$, with $ r_{e}$
the half light radius in kpc and $\tilde{\rho}_e$ the stellar density
in units of solar masses per cubic parsec. Luminosity profiles for $r<r_e$ are well
approximated as power laws with  $1<\gamma<2$ in the same galaxies (Terz\'{i}c \& Graham~2005).
Thus,
\begin{equation}
\Delta t  \approx 3  \times 10^{10} {\rm yr}~~ \tilde{r}_e^{1.75}  \tilde{m}^{-1}~.
\end{equation}
For globular clusters of mass $m=10^5(10^6)M_{\odot}$, $\Delta t<10^{10}~$yr
requires $r_e<140(520)~$pc.
For comparison bulges of S0-Sb galaxies and early type galaxies with  $-20<M_B<-16$ have $200\lesssim r_e \lesssim 5000~$pc , albeit with a large
scatter \citep{F97,F:06,GW08}. Though crude, this calculation   implies  that a significant fraction of the globular clusters  in  faint galaxies and bulges would have spiraled to the center in $10^{10}~$yr.

The  dynamical  evolution  of the Galactic  globular cluster system has been already
investigated by many authors using different schemes and approximations~\citep[e.g.,][]{MW:97,TP00,SKT}. 
None of these previous calculations were however designed to track the cluster orbits down to the sphere of influence of the central black hole. 
A more accurate estimate of orbital inspiral times for clusters in the Galactic bulge is obtained in what follows
by adopting a model  for the background distribution that reproduces more accurately the observed stellar density profile in the inner $\sim 1{\rm kpc}$ of the Milky Way.

We numerically integrated the equations of motion of a cluster in a fixed potential including the contribution of dynamical friction:
\begin{subequations}
\begin{eqnarray}\label{em}
\ddot{\mathbf{r}}&=&- \mathbf{\nabla} \phi +\mathbf{a}_{\rm df}, \\
\mathbf{a}_{\rm df} &=& -4\pi G^2  m \rho(r)    F(<v,r)~
{\rm ln}\Lambda\frac{\mathbf{v}}{v^3}.
\label{dfa2}
\end{eqnarray}
\end{subequations}
Here $\rho(r)$ is the mass density of background stars, $F(<v,r)$ is the fraction of stars 
with local velocities less than that of the cluster and ${\rm ln}\Lambda$ is the Coulomb logarithm.
For the background distribution we adopted the density model:
\begin{eqnarray}
\rho(r)&=&\rho_{\rm NSD}+\rho_{\rm GB}=400~ \frac{M_{\odot}}{{\rm pc^3}}\left( \frac{r}{10{\rm pc}}  \right)^{-0.5}{\rm exp}\left[ -\frac{r}{70 {\rm pc}} \right]\nonumber \\
&&+ 6~ \frac{M_{\odot}}{{\rm pc^3}}{\rm exp}\left[-\left(\frac{r}{800 {\rm pc}}\right)^{2} \right]~, 
\end{eqnarray}
where $\rho_{\rm NSD}$ is the density of the nuclear stellar disk and the
second term, $\rho_{\rm GB}$, represents the contribution of the Galactic bulge which becomes 
significant for $r\gtrsim 200~$pc. This density profile approximates  the  mass model shown in  Figure~14 of \citet{LZM}. 
We chose $m=4\times10^6 M_{\odot}$ for the untruncated mass of the globular cluster. In order to include the effect of tidal truncation, at any time 
the corresponding  tidally-truncated mass ($m_T$), defined in  equation~(\ref{eq:mt}), was computed and  
assigned to the infalling globular cluster (i.e., $m=m_T$) if $m_T<m$.

\begin{figure}
$\begin{array}{cc}\centering
\includegraphics[width=0.4\textwidth,angle=0 ]{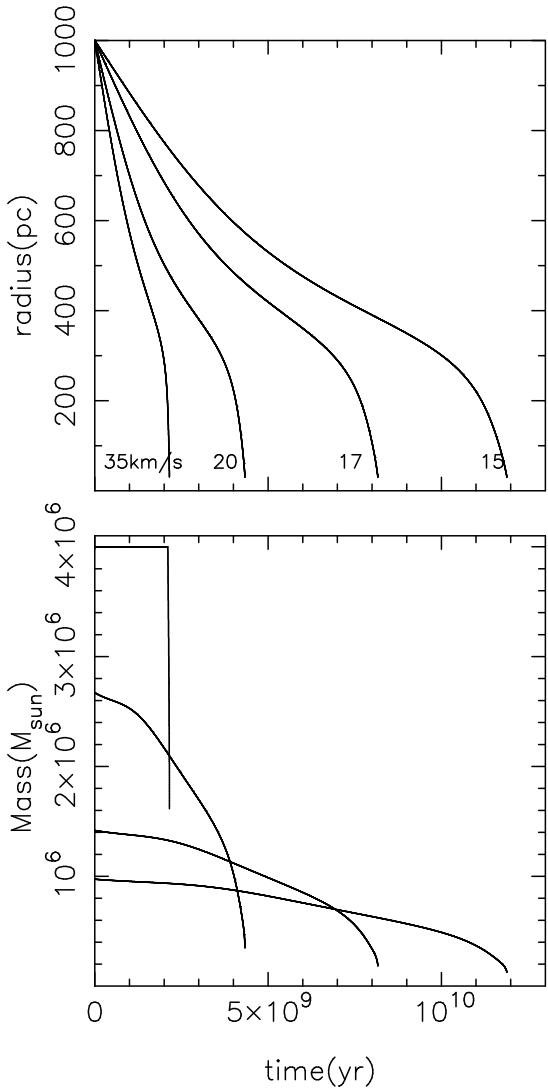} 
\end{array}$
\caption{Evolution of orbital radius (upper panel) and mass (lower panel) 
for globular clusters with different central velocity dispersions starting on circular orbits of radius
$1~$kpc. These computations  include the effect of tidal truncation due to the galactic potential~(see text for explanation).  
Globular clusters with velocity dispersion larger than approximately $15~{\rm km s^{-1}}$ would spiral to the center within $10^{10}~{\rm yr}$.  }\label{decay}
\end{figure}

Figure~\ref{decay}  plots the orbital evolution of globular clusters with different values of the central velocity dispersion and starting on
circular orbits of radius $1~$kpc. 
Globular clusters with central velocity dispersion larger than about $15~{\rm km s^{-1}}$ would reach the center in less
than a Hubble time if they are initially inside the Galactic bulge. 

We stress that the calculations presented in  Figure~\ref{decay}  give a conservative
upper limit to the decay times for various reasons. Some of these are:
i) the presence of gas in the central regions could dramatically decrease the sinking time of a stellar cluster~\citep{Os99};
ii) the background distribution is
assumed to be spherical, but in a more realistic triaxial model  the decay time could be greatly reduced \citep{PE:92}.

In addition, the sinking times given in  Figure~\ref{decay} are a clear overestimate of the real dynamical friction decay times 
since they were evaluated 
considering the Chandrasekhar local approximation for tidally truncated globular clusters on circular orbits (orbits that minimize the dynamical friction effect for a given orbital energy). More accurate orbit integrations, apt to follow the entire evolution of an orbit through the singularity, show that globular clusters on radial orbits decay in a time more than 30 times faster than on circular orbits of same energy \citep{CDAS:11}.
All this convince us that the decay toward the MW central region of sufficiently massive GCs 
has occurred on a time significantly shorter than the age of the Galaxy.

\begin{figure*}
$\begin{array}{ll}
~~~~~~\includegraphics[width=.93\textwidth,angle=0 ]{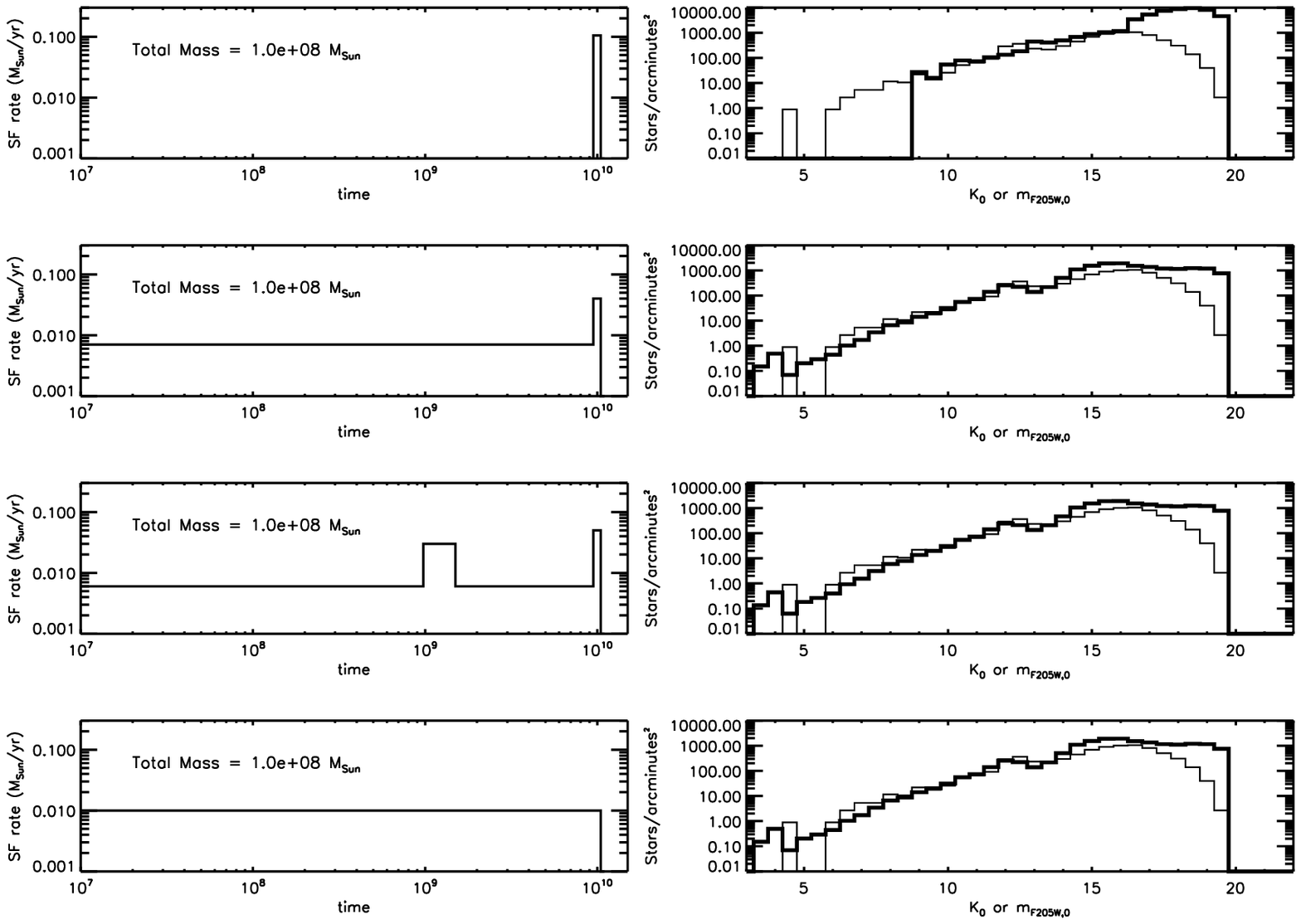}  
\end{array}$
\caption{Left panels display different star formation scenarios in which, from top to bottom:
the entire mass is build  up by ancient ($\sim 10~$Gyr old) stars, that we assume to be brought
into the Galactic center  by infalling globular clusters;
half of the mass is  contributed by old stars and half due to continuous star formation; at the mass contributed
by continuous star formation and globular clusters  we add stars formed during a starburst episode occurred
at $\sim 1~$Gyr;  all the mass is contributed by continuous star formation.
Right panels compare the observed LF of the Galactic center  (light lines) with the
LFs (heavy lines) resulting from the different star formation scenarios assuming Solar metallicity and  canonical
mass loss-rates in the Geneva models.  
Note  that the data are much more than $50\%$ incomplete for the faintest few bins.
The models have not been scaled for mass, but rather have been rescaled along the vertical axis
   to mach the number counts in the $K=11.0$ bin.  
The star formation histories corresponding to the three bottom set of panels are essentially indistinguishable from each other and they
all reproduce quite well the observed LF.}\label{sfr}
\end{figure*}

\subsection{Star Formation History of the Milky Way Nuclear Cluster}
The NSC at the center of the Milky Way appears to have undergone
continuous  star formation over the last ~$10~$Gyr~\citep[e.g.,][]{SM96}.
The  evidence that a large fraction of the Galactic center mass  was formed in situ at an approximately  constant rate over the last $10~$Gyr
has been extensively used in the past to argue against the merger model for the Milky Way NSC~\citep[e.g.,][]{Milos04,Nay09} .

In this section, we compare the observed luminosity function~(LF) of the stellar populations at the Galactic center with
synthetic LFs obtained for different star formation scenarios.  Our analysis suggests
the possibility that about   $1/2$ of the Milky Way nuclear  population consists of  $\sim 10~$Gyr old stars
brought  in by  infalling globular clusters, while  the remaining mass  is due to continuous star formation.

Dereddened LFs of the observed populations were
generated  by  {\it Hubble Space Telescope} NICMOS data taken from \citet{Figer2004}. The fields are all within the central $30~$pc
of the Galactic center; their exact locations are
given in Table~1 and shown in Figure~1 of that paper.
Data are complete at the $50\%$ level
at $m_{\rm F205W}=19.3$, averaged over all fields.
The details of the technique used to generate both model and observed
LFs are also described in \citet{Figer2004} and are summarized in what follows.  

The LFs of the Galactic center populations are constructed under the assumption that each star has the intrinsic colors of a red giant and by
subtracting reddening values for each star corresponding to its color in $H-K$ or $m_{\rm F160W}-m_{\rm F205W}$.

Geneva isochrones are used to model the LFs corresponding to different star formation histories.  
The Geneva stellar evolutionary models are described in  \citet{Sh92}, \citet{S93a, S93b},
\citet{Ch93} and \citet{Me94}.
We adopted  a mass spectrum of stars with a Salpeter \citep{salpeter} power law index  (i.e., $dN/dm=m^{-1.35}$) and upper and lower mass
cutoff  of $120~M_{\odot}$ and $0.1~M_{\odot}$ respectively.
The Geneva models are used to convert these masses to the absolute magnitudes in the $V$ band that
are subsequently  transformed in the $K$ band through a lookup table  that relates color index to magnitude.
We then sum over the histogram to produce the LF of each star formation event and sum the individual LFs
to derive the  LF for a given star formation scenario.

Figure~\ref{sfr} displays the results of this study for various star formation histories, with the model
counts modified by the observed completeness  fractions from \citet{F99}.  
From  the  top to the bottom panels, the plots  correspond to   star formation models  in which: i) the entire mass is build
up by ancient ($\sim 10~$Gyr old) stars, possibly brought into the Galactic center  by infalling globular clusters; ii)  the mass model
is composed by ancient  globular clusters stars plus stars formed via continuous star formation;  iii) some ($\sim 1/3$) of the mass is formed during a starburst
at $\sim 1~$Gyr  \citep[e.g.,][]{Sj}, while  the remainder is due to continuous star formation and to  
an ancient burst at  $10~$Gyr; iv) the entire mass is formed through continuous star formation  over the last
$10~$Gyr.

The counts at faint magnitudes ($K_0>15$) are controlled by ancient star formation, while the
counts at the bright end ($K_0 <8 $) are  controlled by the extent of recent star formation activity.
The brightness of the red clump (at $K_0 \sim 12$) is
related instead to the extent of intermediate age star formation activity.

The figure shows that the ancient burst model, corresponding to a NSC composed of only ancient stars,
fails at reproducing the observed LF. This model overestimates  the counts at  faint magnitudes and it does not
reproduce the number of counts seen in the bright end. Our analysis rules out the possibility that
the nuclear population consists entirely of ancient stars.
On the other hand, star formation  models  in which ancient bursts are
accompanied by continuous star formation at other times, produce a LF essentially indistinguishable
from that  obtained when  the mass is entirely due to continuous star formation.
All these latter models fit quite well  number counts and shape of the observed LF.
We conclude that the observed Galactic center  LF  is consistent
with a star formation history in which a large fraction of the mass consists of ancient~( $\sim10~$Gyr) stars.
A similar result was obtained recently by \citet{Ol11}. These authors 
constructed  a  complete Hertzprung Russel diagram of the red giant population within $1~$pc
from Sgr A* and found  that about  $80\%$ of the stellar mass in these regions 
was formed more than $5~$Gyr ago.

\subsection{Mass-radius relation}
\begin{figure}
\centering
\includegraphics[width=0.4\textwidth,angle=270 ]{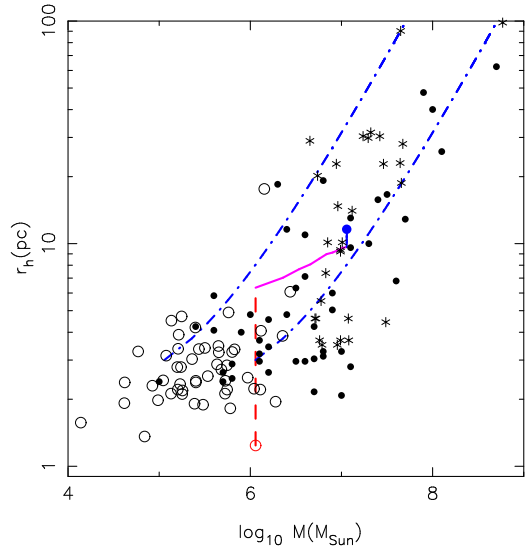}
\caption{The measured mean half-mass radius (or effective radius)   plotted against total
mass for nuclei (filled circles), globular clusters (empty circles) and UCDs (stars symbols).
Data points are from \citet{FORBES} and \citet{Cote}.
Blue dot-dashed curves show the predicted scaling
in the merger model without \sbh\ for two different choices for the mass of merged clusters (see text
for details). The red-open circle represents the initial globular cluster model
in the $N$-body simulation. Purple and continue-blue curves are
the evolutionary track of the NSC during its formation and during relaxation respectively.
The filled blue circle represents the final product of our simulation.}\label{obs}
\end{figure}

In Figure~\ref{obs}  the  mean half-mass radius  is plotted against total
mass for nuclei (filled circles) globular clusters (open circles) and ultra compact dwarfs (UCDs, star symbols).
We overplot the track followed by the NSC in our simulation during the infall events (purple-continue curve)
and during relaxation (continue-blue line).
The  structural properties   of the NSC formed  in our simulations (blue-filled circle) are in  good agreement with those of real NSCs.

From the figure we can see that  the faintest nuclei
have roughly the same mass as a typical globular cluster. The size distributions for the nuclei
and globular clusters also overlap, although the clusters in the Galaxy  have half-mass radii of $3~$pc, irrespective of mass,
while the nuclei follow a relation of the form $r_h\propto \sqrt{M}$.
Fainter than a few million solar masses, the nuclei and globular clusters have comparable sizes \citep{Has05}.

We now consider the merger model for nucleus formation in the absence of a \sbh.
In this case one can derive a simple recursive relation between the mass and radius of the NSC during its formation.
The radius of the nucleus increases with increasing total mass, or light, as globular clusters
merge. After the merger, its final energy, $E_f$ , equals the energy of the nucleus before the merger, $E_i$,
plus the energy brought in by the globular cluster. This energy has two components: the internal
energy or binding energy $E_b$, and the orbital energy just before the merger, $E_o$. From conservation
of energy:
\begin{equation}
E_f=E_i+E_o+E_b~.
\end{equation}
Just before the merger, the orbital energy is $E_o=\alpha G m M_{i}/2R_{i}$, where
$M_{i}$  and $R_{i}$ are the
mass and radius of the nucleus, respectively, $m$ is the mass of the globular cluster, and
$\alpha$ is a constant of order unity \citep{Hausman} that depends on the radius of
the capture orbit - the radius at which the
dominant influence on the trajectory of a globular cluster first comes from the nucleus. After the
merger, the nucleus reaches a state of dynamical equilibrium quickly; the virial theorem implies
$E_f=-G M_{f}^2$.
The equations above permit expressing the mass, energy, and radius of the
nucleus recursively as

\begin{eqnarray}
M_{j+1}=(j+1)M_{1},~~~~~~~~~~~~~~~~~~~~~~ \label{rec1}\\
jE_{j+1}=\left( j+\alpha \right)E_{j}+jE_1,~~~~~~~~~~~~~~~~~~~\label{rec2}\\
\left( j+1\right)^2 R^{-1}_{j+1}=j\left( j+\alpha \right)R^{-1}_j +R^{-1}_1, j=1,2,3,... ~~~~~~~~\label{rec3}
\end{eqnarray}
where the subscript 1 denotes the initial nucleus, and, by assumption, $M_1=m$.
At the time when a nucleus consists of few merged globular clusters, its mass and that of
the next infalling globular cluster are comparable. In this case, equations~(\ref{rec1},\ref{rec3}) imply $R\propto M^{0.5}$.
However, after many mergers, $M\gg m$, and the relation steepens to $R\propto M$. For $\alpha=1.2$ and
5(25)100 mergers, equations~(\ref{rec1},\ref{rec3})  imply $R\approx 2(5)10$ and $R\propto M^{p}$, $p=0.5(0.6)0.7$. The
typical half-mass radius of a globular cluster is about 3 pc \citep{Jord} so for a nucleus assembled from 25
mergers, $R\sim 15$ pc. This is in reasonable agreement with the measured sizes for the brighter
nuclei. For $\alpha=1.2$, the expected scaling between $r_h$ and mass is shown by the blue dot-dashed
curves  in  Figure~\ref{obs}. We show the predicted behavior for two assumptions for the
mass, $m$, of the clusters which merge to form the nucleus: $10^5$
and $10^6~M_{\odot}$.
At least for $m=10^6~M_{\odot}$, the agreement with the $r_h$-mass relation for real nuclei is remarkably good.

\section{Summary}
We have used large-scale direct $N$-body simulations to test the merger model for the
formation of the Milky Way nuclear star cluster (NSC).
Our initial conditions consisted of a massive black hole (\sbh) at the center of
a nearly homogeneous $N$-body system representing the nuclear stellar disk.
Globular clusters were then added to this system, starting from circular orbits
of radius 20 pc. The clusters were tidally limited by the external field to have
a mass of $\sim 1.1\times 10^6$ M$_\odot$ at the start.
Infall was driven by dynamical friction, due to the stellar disk, and later also
to the accumulated mass from the previously-merged clusters.
The clusters were fully disrupted by the \sbh\ at a radius of approximately
one parsec. After 12 inspiral events, the accumulated mass of the NSC
was about $1.5\times 10^7M_\odot$, comparable with the actual mass.

The principle results of our study are summarized below.

\begin{itemize}
\item[1] The stellar system resulting from the consecutive mergers has a density
that falls off as $\sim r^{-2}$, and a core of radius $\sim 1$ pc. These
properties are
similar to those observed in the Milky Way NSC.
\item[2] The morphology of the NSC evolved during the course of the infalls,
from
an early, strongly triaxial shape toward a more oblate/axisymmetric shape
near the end of the merger process.
Kinematically, the final system is characterized by  a mild tangential
anisotropy
within the inner $30~$pc and a low degree of rotation.
\item[3] In order to investigate the effect of gravitational encounters on the
evolution of the NSC, we continued the $N$-body integrations after the final
inspiral was complete.
The core that had been created by the \sbh\ was observed to shrink by roughly a
factor of two in 10 Gyr as the stellar density evolved toward a Bahcall-Wolf
cusp.
This final core size is essentially identical to the size of the core observed
at the
center of the Milky Way.
The density profile outside the core remained nearly unchanged during this
evolution.
Gravitational encounters also caused the NSC to evolve toward spherical symmetry
in configuration and velocity space.
\item[4] 
Since Galactic globular clusters are almost exclusively ancient objects with ages $\sim 10-13~$Gyr, 
in the merger model  a large fraction of  the NSC
 mass is expected to be in old stars.
We confronted this prediction with the observed luminosity function (LF)  of the
Milky Way NSC.
Using stellar population models, we showed that the observed LF  is consistent
with a star formation history in which a large fraction (about 1/2) of  the mass
consists of old ( $\sim10~$Gyr) stars and the remainder from continuous star
formation.
\end{itemize}

\bigskip

This research was supported by the National Science Foundation under
grants  AST 08-07910, 08-21141 and by the National Aeronautics and
Space Administration under grant NNX-07AH15G.
We are grateful to Don Figer that provided us with  data and code used to generate Figure~\ref{sfr}.
We thank H.~Perets, C.~Trombley , E.~Vasiliev  for useful discussions and the referee,
S.~Portegies Zwart, for comments that helped to improve the paper.

\end{document}